\documentclass[sigconf]{acmart}

\renewcommand\footnotetextcopyrightpermission[1]{} 
\AtBeginDocument{%
  }

\usepackage{import}
\newcounter{rqcounter}
\newcommand{\newrq}[2]{\noindent\refstepcounter{rqcounter}\textbf{RQ\arabic{rqcounter}:} {\em #2}\label{#1}}
\newcommand{\rqref}[1]{\textbf{RQ\ref{#1}}}
\newcommand\alias[2]{\expandafter\def\csname alias:#1\endcsname{#2}}
\newcommand\A[1]{\csname alias:#1\endcsname}

\usepackage{enumitem}

\definecolor{mygreen}{RGB}{189,178,70}
\definecolor{mypink}{RGB}{202,45,85}
\definecolor{myblue}{RGB}{0,57,92}
\definecolor{highlightgreen}{rgb}{0.0, 0.5, 0.0}  %

\newcommand\Revision[1]{\textcolor{black}{#1}}
\newcommand\MandatoryRevision[1]{\textcolor{black}{#1}}

\usepackage{fdsymbol}
\usepackage{booktabs}
\usepackage{tabularx}
\usepackage{multirow}
\usepackage{comment}
\usepackage{graphicx}
\usepackage{url}
\usepackage{circledsteps}
\usepackage{tcolorbox}
\usepackage{caption}
\usepackage{subcaption}
\usepackage{colortbl}

\usepackage{longtable}
\usepackage{wasysym}

\tcbset{
  myhighlightbox/.style={
    colback=myblue!5!white,
    colframe=myblue!75!black,
    boxrule=0.8pt,
    arc=2pt,
    left=4pt,
    right=4pt,
    top=4pt,
    bottom=4pt
  }
}

\usepackage[%
acronym
]{glossaries}

\newglossaryentry{log4j}{%
    name={Log4j},
    description={},
}
\newglossaryentry{solarwinds}{
    name={SolarWinds},
    description={},
}
\newglossaryentry{xz}{
    name={XZ Utils},
    description={},
}
\newglossaryentry{attack}{
    name={attack technique},
    description={},
    first={attack techniques}
}
\newglossaryentry{safeguard}{
    name={task},
    description={},
    first={tasks},
}
\newglossaryentry{starter}{
    name={\textit{starter kit}},
    description={},
}

\newacronym{cncf}{CNCF}{Cloud Native Computing Foundation}
\newacronym[first=Secure Software Development Framework (SSDF)]{ssdf}{SSDF}{Secure Software Development Framework}
\newacronym{psscrm}{P-SSCRM}{Proactive Software Supply Chain Risk Management}
\newacronym{slsa}{SLSA}{Supply-chain Levels for Software Artifacts}
\newacronym{eo}{EO}{Executive Order}
\newacronym{scorecard}{OpenSSF Scorecard}{OpenSSF Scorecard}
\newacronym{bsimm}{BSIMM}{Building Security In Maturity Model}
\newacronym{s2c2f}{S2C2F}{OpenSSF Secure Supply Chain Consumption Framework}
\newacronym{owaspSCVS}{OWASP SCVS}{OWASP Software Component Verification Standard}
\newacronym{cncfSSC}{CNCF SSC}{CNCF Software Supply Chain Best Practices}
\newacronym{attestation}{Self-Attestation}{Self-Attestation}
\newacronym{attck}{MITRE ATT\&CK}{MITRE Adversarial Tactics, Techniques and Common Knowledge}
\newacronym[first={Cyber Threat Intelligence (CTI)}]{cti}{CTI}{Cyber Threat Intelligence}
\newacronym{nist}{NIST}{National Institute of Standards and Technology}
\newacronym[firstplural=Common Vulnerabilities and Exposures (CVEs)]{cve}{CVE}{Common Vulnerabilities and Exposures}
\newacronym{cwe}{CWE}{Common Weakness Enumeration}
\newacronym{cisa}{CISA}{Cybersecurity and Infrastructure Security Agency}
\newacronym{nvd}{NVD}{National Vulnerability Database}
\newacronym{cna}{CNA}{CVE Numbering Authority}
\newacronym{llm}{LLM}{Large Language Models}
\newacronym[first=Software Bills of Materials (SBOMs)]{sbom}{SBOM}{Software Bill of Materials}
\newacronym{ttps}{TTPs}{Tactics, Techniques, and Procedures}
\newacronym{cot}{CoT}{Chain of Thought}
\newacronym{jndi}{JNDI}{Java Naming and Directory Interface}
\newacronym{poc}{PoC}{Proofs of Concept}
\newacronym{tp}{TP}{True Positive}
\newacronym{fp}{FP}{False Positive}
\newacronym{fn}{FN}{False Negative}
\newacronym[first=Attacks Under Study (AUS)]{aus}{AUS}{Attacks Under Study}

\newacronym{2fa}{2FA}{two-factor authentication}
\newacronym{bdfl}{BDFL}{Benevolent Dictator for Life}
\newacronym{capec}{CAPEC}{Common Attack Pattern Enumeration and Classification}
\newacronym{cra}{CRA}{Cyber Resilience Act}
\newacronym{ci}{CI}{continuous integration}
\newacronym{cicd}{CI/CD}{continuous integration and continuous delivery}
\newacronym{cipac}{CIPAC}{Critical Infrastructure Partnership Advisory Council}
\newacronym{cla}{CLA}{Contributor License Agreement}
\newacronym{cvss}{CVSS}{Common Vulnerability Scoring System}
\newacronym{dpo}{DPO}{Data Protection Officer}
\newacronym{enisa}{ENISA}{European Union Agency for Cybersecurity}
\newacronym{epss}{EPSS}{Exploit Prediction Scoring System}
\newacronym{esf}{ESF}{Enduring Security Framework}
\newacronym{gdpr}{GDPR}{General Data Protection Regulation}
\newacronym{irb}{IRB}{Institutional Review Board}
\newacronym{ngo}{NGO}{non-governmental organization}
\newacronym{ntia}{NTIA}{National Telecommunications and Information Administration}
\newacronym{openssf}{OpenSSF}{Open Source Security Foundation}
\newacronym{oss}{OSS}{open source software}
\newacronym{osc}{OSC}{open source component}
\newacronym{osp}{OSP}{open source project}
\newacronym{sat}{SAT}{static analysis tool}
\newacronym{satc}{SaTC}{Secure and Trustworthy Cyberspace}
\newacronym{sme}{SME}{small and medium enterprises}
\newacronym{ssc}{SSC}{software supply chain}
\newacronym{sso}{SSO}{single sign-on}
\newacronym{svm}{SVM}{Support Vector Machine}
\newacronym{tls}{TLS}{Transport Layer Security}
\newacronym{vcc}{VCC}{vulnerability-contributing commit}
\newacronym{vex}{VEX}{Vulnerability-Exploitability Exchange}

\setcopyright{acmlicensed}
\copyrightyear{2025}
\acmYear{2025}
\acmDOI{XXXXXXX.XXXXXXX}
\acmConference[Conference acronym 'XX]{Make sure to enter the correct
  conference title from your rights confirmation email}{June 03--05,
  2018}{Woodstock, NY}

\acmISBN{978-1-4503-XXXX-X/18/06}

\begin{document}

\title{Closing the Chain: How to reduce your risk of being SolarWinds, Log4j, or XZ Utils}

\newcommand{\ncsuimg}{\raisebox{-0.3ex}{\includegraphics[height=1.8ex]{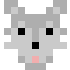}}}
\newcommand{\yahooimg}{\raisebox{-0.3ex}{\includegraphics[height=1.8ex]{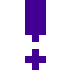}}}
\newcommand{\shorthillimg}{\raisebox{-0.3ex}{\includegraphics[height=1.8ex]{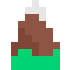}}}

\author{Sivana Hamer\textsuperscript{\ncsuimg}, Jacob Bowen\textsuperscript{\ncsuimg}, Md Nazmul Haque\textsuperscript{\ncsuimg}, Chris Madden\textsuperscript{\yahooimg}, Robert Hines\textsuperscript{\shorthillimg}, Laurie Williams\textsuperscript{\ncsuimg}}

\affiliation{%
  \institution{\textsuperscript{\ncsuimg}North Carolina State University, \textsuperscript{\yahooimg}Yahoo Inc., \textsuperscript{\shorthillimg}Short Hill Advancements}
  \country{}
}
\email{{sahamer,jebowen2,mhaque4,lawilli3}@ncsu.edu, chris.madden@yahooinc.com, rmhines@shorthilladvancements.com}

\renewcommand{\shortauthors}{Hamer et al.}

\begin{abstract}

\Revision{Software supply chain frameworks, such as the US NIST \gls{ssdf}, detail \textit{what tasks} software development organizations \Revision{are recommended or mandated to} adopt to reduce security risk.}
\MandatoryRevision{However, to further reduce the risk of similar attacks occurring, software organizations benefit from knowing \textit{what tasks mitigate attack techniques} the attackers are currently using to address specific threats, prioritize tasks, and close mitigation gaps.}
\MandatoryRevision{\textbf{
The goal of this study is
to aid software organizations
in reducing the risk of software supply chain attacks
by systematically synthesizing how framework tasks mitigate the \glsfirst{attack} used in the \gls{solarwinds}, \gls{log4j}, and \gls{xz} attacks.
}}
We qualitatively analyzed 106 \gls{cti} reports of the 3 attacks to gather the attack techniques.
We then systematically constructed a mapping between attack techniques and the 73 tasks enumerated in 10 software supply chain frameworks.
Afterward, we established and ranked priority tasks that mitigate attack techniques.
The three mitigation tasks with the highest scores are role-based access control, system monitoring, and boundary protection.
Additionally, three mitigation tasks were missing from all ten frameworks, including sustainable open-source software and environmental scanning tools.
Thus, software products would still be vulnerable to software supply chain attacks even if organizations adopted all recommended tasks.
\end{abstract}



\received{20 February 2007}
\received[revised]{12 March 2009}
\received[accepted]{5 June 2009}

\maketitle

\section{Introduction}

In software supply chain attacks, vulnerabilities are maliciously or accidentally introduced into dependencies, compromising direct or transitive dependents.
Attackers exploit these software dependencies, the build infrastructure, and humans involved as attack vectors~\cite{ladisa2023sok, williams2024research}.
While software supply chain attacks are not new, since 2020, high-profile incidents—such as those affecting \gls{solarwinds}, \gls{log4j}, and \gls{xz}—have been discovered, compromising thousands of government, industry, and open-source organizations worldwide~\cite{cisa2021Joint, csrb2022Log4j, cox24Timeline}.
Trends, such as the annual detected malicious open-source packages rising from 929 in 2020 to 459,070 in 2024~\cite{sonatype2021, sonatype2024}, 
indicate the growth of software supply chain attacks.

\Revision{Tasks have been proposed and researched to reduce the risk of software supply chain attacks, including \gls{sbom} creation, code signing, and reproducible builds.}
Government and industry have developed software supply chain risk management frameworks, compiling \textit{what tasks software organizations should adopt} to reduce security risk.
Notable frameworks include: US \gls{nist} \glsfirst{ssdf}~\cite{ssdf}, \acrlong{scorecard}~\cite{scorecardOpenSSF}, and
\gls{slsa}~\cite{slsa}.
\Revision{Regulations such as US \gls{eo} 14028~\cite{eo14028}, US \gls{eo} 14144~\cite{eo14144,amendmentEO14144}, and the European Cyber Resilience Act (CRA)~\cite{cra} mandate adopting software supply chain tasks.}

\MandatoryRevision{
Despite increased framework adoption~\cite{tidelift2024}, little work has evaluated software supply chain framework tasks against the mitigated threats.
First, although work has evaluated threats through models and attack vectors~\cite{slsa,s2c2f,ladisa2023sok}, conducting risk assessment through attack techniques enables addressing specific threats by understanding what adversaries desire and do~\cite {nist80030}.
Hence, framework tasks can be selected to mitigate specific risks of attacks.
Second, the \gls{psscrm}~\cite{williams2024proactive} framework unified 10 prominent software supply chain frameworks and found 73 tasks.
Given that organizations have limited budget, time, and resources for software supply chain security~\cite{sammak2023developers}, adopting all 73 tasks is impractical. 
Thus, organizations benefit from a ranked priority to focus on essential mitigation tasks.
Finally, without evaluating whether gaps exist for the framework tasks based on the attack techniques, software products could still be vulnerable to attacks.
}
\MandatoryRevision{\textbf{
The goal of this study is
to aid software organizations
in reducing the risk of software supply chain attacks
by systematically synthesizing how framework tasks mitigate the \glsfirst{attack} used in the \gls{solarwinds}, \gls{log4j}, and \gls{xz} attacks.
}}
\Revision{We address the following research questions:}

\alias{rq:events-text}{What are the \glsfirst{attack} used in the \gls{solarwinds}, \gls{log4j}, and \gls{xz} attacks?}
\newrq{rq:events}{\A{rq:events-text}}

\alias{rq:safeguards-text}{What software supply chain framework \glsfirst{safeguard} mitigate \glsfirst{attack} in the \gls{solarwinds}, \gls{log4j}, and \gls{xz} attacks?}
\newrq{rq:safeguards}{\A{rq:safeguards-text}}

\alias{rq:frameworks-text}{Which software supply chain mitigation \glsfirst{safeguard} are missing from the frameworks based on the \gls{solarwinds}, \gls{log4j}, and \gls{xz} attacks?}
\newrq{rq:frameworks}{\A{rq:frameworks-text}}

\MandatoryRevision{To that end, we meta-synthesized~\cite{ralph2022paving} the \gls{solarwinds}, \gls{log4j}, and \gls{xz} attacks, which we call \gls{aus}.} 
\Revision{We study~\cite{mann2003observational} the \gls{aus} through 106 reports and collect adversarial events.}
We then mapped the adversarial events to attack techniques, detailing how attacks were carried out, using \gls{attck}~\cite{mitreATTACK}.
\Revision{We then map the attack techniques to \gls{psscrm} tasks using four strategies.}
\Revision{Afterward, we identify gaps in \gls{psscrm} (and transitively in the \Revision{ten} reference frameworks) by qualitatively analyzing the \gls{cti} reports.}
\MandatoryRevision{Finally, we constructed an evidence-based listing of the most important tasks to adopt to mitigate current attack techniques in a ranked \gls{starter} mapped to software supply chain frameworks with task and framework measures.}

\Revision{We found that the 3 \gls{aus} have a common set of 12 attack techniques leveraged by adversaries, including exploiting trusted relationships, obfuscating data, and compromising infrastructure.}
\Revision{Only 34 out of 73 tasks in \gls{psscrm} mitigate \gls{aus} attack techniques.}
The \gls{starter} tasks with the top three highest scores are role-based access control (E.3.3), system monitoring (D.2.1), and boundary protection (E.3.7).
Finally, we found three missing tasks in \gls{psscrm} and, therefore, in all ten contributing frameworks. 
Hence, products of software organizations would still be vulnerable to attacks even if adopting all tasks in the 10 frameworks referenced in \gls{psscrm}.

\MandatoryRevision{In summary, we contribute:
\Circled{1} a systematic empirical procedure with measures to model how tasks mitigate attack techniques in software security attacks;
\Circled{2} a prioritized evidence-based list of \gls{starter} tasks for software organizations to adopt for software supply chain attacks;
\Circled{3} a list of mitigation tasks gaps in software supply chain frameworks from \gls{cti} reports;
\Circled{4} event timelines of \gls{solarwinds}, \gls{log4j}, and \gls{xz} mapped to attack techniques with associated threat models; and,
\Circled{5} triangulated \gls{attck} attack techniques to \gls{psscrm} tasks mappings.}

The remainder of this paper is as follows.
\Revision{First, we detail our methodology (Sec.~\ref{sec:metho}).
After, we present our results (Sec.~\ref{SEC:RQ1}, \ref{SEC:RQ2}, and \ref{SEC:RQ3}).
We then discuss our findings (Sec.~\ref{sec:discussion}) and \Revision{limitations} (Sec.~\ref{sec:threats}).
We finalize by describing related work (Sec.~\ref{sec:relwork}) and concluding (Sec.~\ref{sec:conc}).}

\begin{figure}[t]
    \centering
    \includegraphics[width=\linewidth]{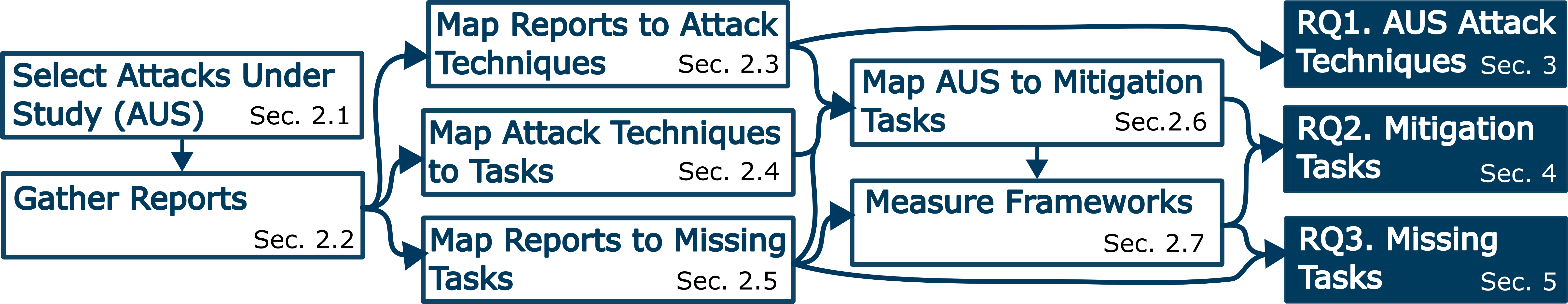}
    \caption{\Revision{Research methodology flow chart}}
    \label{fig:Methodology}
    \Description[Research Methodology Flow Chart with the Relationship to the Research Questions]{}
\end{figure}

\section{Methodology}
\label{sec:metho}

\MandatoryRevision{In this section, we describe our meta-synthesis methodology that empirically constructs a model of how tasks mitigate attack techniques in software supply chain attacks.
Meta-synthesis~\cite{ralph2022paving} is a secondary research method that summarizes qualitative data through qualitative methods.
For example, we synthesized \gls{cti} reports (i.e., secondary data) with thematic analysis~\cite{cruzes2011recommended} (i.e., qualitative method).}
\MandatoryRevision{
To increase transparency, we detail the authors' expertise in a \textbf{statement of positionality}.
Six authors were involved in the research: two practitioners with more than two decades of experience in security; one who taught and researched security for two decades, publishing in security conferences; one who has taken graduate-level courses in security and conducted research for two years; and two advised by the other authors.}

\MandatoryRevision{
A methodology overview is shown in Fig.~\ref{fig:Methodology}.
Our process draws upon incident analysis, a process to prevent future incidents by identifying contributing factors~\cite{leveson2016engineering}, applied in incidents such as Therac-25~\cite{leveson1993investigation}.
Incident analysis approaches define the scope, collect incident data, analyze incidents, determine causes, report results, and provide recommendations to industry~\cite{amusuo2022reflections}.
In line, we start by selecting the \gls{aus} (Sec.~\ref{SEC:Metho-Sel})} \MandatoryRevision{and gathering \gls{cti} reports (Sec.~\ref{SEC:Metho-Col}). 
We then map reports to attack techniques (Sec.~\ref{SEC:Metho-Qual}), construct an attack technique to task mapping (Sec.~\ref{SEC:Metho-Map}), and identify frameworks gaps (Sec.~\ref{SEC:Metho-Gap}).}
\MandatoryRevision{Finally, we mapped the \gls{aus} to mitigation tasks (Sec.~\ref{SEC:Metho-Val}), and measured the frameworks (Sec.~\ref{SEC:Metho-Measure}).}

\subsection{Select Attacks Under Study (AUS)}\label{SEC:Metho-Sel}

\Revision{We define software supply chain attacks as incidents where vulnerabilities in dependencies are leveraged by adversaries in dependent systems~\footnote{\Revision{We broaden prior definitions (e.g., ~\cite{ladisa2023sok, okafor2022sok, ohm2020backstabber}) as accidentally vulnerable components (e.g., \gls{log4j}) are also considered as software supply chain compromises by industry~\cite{sonatype2024}.}}. 
We reviewed the Cloud Native Computing Foundation (CNCF) catalog of compromises~\cite{catalogSSC} and searched for attacks. %
We purposefully selected diverse and notable attacks in line with case studies~\cite{runeson2009guidelines}.}
\Revision{We selected three cases as the \gls{aus}: \gls{solarwinds}, \gls{log4j}, and \gls{xz}.}
\Revision{First, the main initial attack vector~\cite{williams2024research} varied between the \gls{aus}, targeting the build infrastructure (\gls{solarwinds}), dependencies (\gls{log4j}), and humans (\gls{xz}).}
\Revision{Second, the attacks garnered a high degree of attention from the community due to the sophistication or the number of compromised organizations.}

\subsection{Gather Reports}\label{SEC:Metho-Col}

\MandatoryRevision{
We use Cyber Threat Intelligence (CTI) reports to retrospectively study~\cite{mann2003observational} the \gls{aus}.}
\Revision{\gls{cti} reports are compiled by organizations to share threat-related information~\cite{johnson2016guide}.}
\MandatoryRevision{Retrospective case studies are observational research where past cases are analyzed without researcher intervention~\cite{mann2003observational}.}
\Revision{We use \gls{cti} reports for three main reasons.
First, interviewing either the adversaries, in some cases, are nation states and parties involved, which have been overwhelmed by the media, is legally and logistically difficult.
Second, interviewing experts on the attacks would be complex as we would need to: (a) sample a small, hard-to-identify group of experts; and (b) conduct hours-long interviews as the attacks span years.
}
\Revision{Third, reports serve as a data source to analyze software engineering incidents~\cite{anandayuvaraj2024fail} and \gls{cti} reports are a source of threat information~\cite{johnson2016guide}.}
To find reports of interest, we use multiple sampling strategies~\cite{baltes2022sampling} to ensure we collect a diverse set of reports of the \gls{aus} and are not limited to one type of information source.
Our sampling strategy focused on gathering information from expert sources.
As discussed in the next paragraphs, we use three sampling strategies to find reports: from notable organizations (G1), are listed in \gls{nist} \gls{nvd} (G2), and from notable parties (G3).
We then discuss how we select the reports we analyze.

\textbf{Notable organizations (G1):} 
We purposefully search for reports from notable software security organizations interested in software supply chain security. 
We selected four notable organizations to provide a diverse set in aim and scope: 
\gls{cisa}~\cite{cisa}, a US government agency protecting against \Revision{cybersecurity} threats; 
MITRE~\cite{mitre}, a non-profit research and development organization supporting government defense agencies;
OpenSSF~\cite{openssf}, a cross-industry community improving the security of open-source software;
and the Linux Foundation~\cite{linuxfoundation}, a non-profit organization supporting open-source development. 
Two researchers independently searched for reports about the \gls{aus} on each organization's website.
We found 49 candidate reports: 25 of \gls{solarwinds}, 18 of \gls{log4j}, and 6 of \gls{xz}.

\textbf{\gls{nist} \gls{nvd} (G2):}
\MandatoryRevision{
Vulnerability databases store vulnerability data, commonly cataloged through \gls{cve}~\cite{cve}, publicly disclosed vulnerabilities.
We selected \gls{nist} \gls{nvd}~\cite{nistNVD} as the database is commonly used in industry~\cite{anwar2021cleaning}.
}
\Revision{Each \gls{cve} listed in \gls{nist} \Revision{\gls{nvd}}}
contains
public references provided by authorized entities to submit details about the vulnerability~\cite{cveProcess}.
We gathered all listed references in the \glspl{cve} of \gls{solarwinds} (CVE-2020-10148), \gls{log4j} (CVE-2021-44228, CVE-2021-45046, CVE-2021-4104, CVE-2021-45105, CVE-2021-44832), and \gls{xz} (CVE-2024-3094).
We found 179 candidate reports: 2 of \gls{solarwinds}, 123 of \gls{log4j}, and 54 of \gls{xz}.

\textbf{Notable parties (G3):} 
Since thousands of systems were compromised in the \gls{aus}, we focused on parties (organizations or people) targeted for initial access, involved, or discovered the incidents.
The list of notable parties was compiled by two authors after analyzing G1 and G2 (Sec.~\ref{SEC:Metho-Qual}) to have a deeper understanding of the \gls{aus}.
The parties in:
\gls{solarwinds} are FireEye and SolarWinds; 
\gls{log4j} are Apache, Alibaba, and Chen Zhaojun (disclosed vulnerability); and,
\gls{xz} are XZ Utils, LZMA, OSS-Fuzz, libsystemd, sshd, libarchive, Lasse Collins (original maintainer), Andres Freund (disclosed vulnerability), and Sam James (accidentally helped attacker).
The first author manually looked at websites, pages, and repositories of or about the notable parties to find \gls{aus} related reports. 
We found 34 candidate reports: 14 of \gls{solarwinds}, 2 of \gls{log4j}, and 18 of \gls{xz}.

\textbf{Report selection.}
From G1, G2, and G3 we found 262 candidate reports. 
We select reports relevant to our study by applying the following inclusion criteria:
(a) is in English;
(b) is publicly available;
(c) is not behind a paywall; 
(d) is not promotional material;
(e) is unique;
(f) contains \gls{cti};
(g) is related to the specific attack;
(h) contains information about attack events or task recommendations;
and (i) has text.
\Revision{Two researchers independently analyzed each criterion for each candidate article and used negotiated agreement~\cite{campbell2013coding} to discuss and reconcile disagreements.}
We selected 106 reports (29 G1, 55 G2, 22 G3): 30 for \gls{solarwinds}, 45 for \gls{log4j}, and 31 for \gls{xz}.
The full list is available in the supplemental material~\cite{supplementalLink}.

\subsection{\Revision{Map Reports to Attack Techniques}}\label{SEC:Metho-Qual}

\MandatoryRevision{We qualitatively analyzed 106 reports using thematic analysis.
Thematic analysis is a method to synthesize qualitative data to report themes within the data~\cite{cruzes2011recommended}.
We therefore analyzed the 106 reports to understand the chain of events that transpired in the \gls{aus} with associated attack techniques to later map to tasks (Sec.~\ref{SEC:Metho-Map}).
Our method had three phases: select an attack technique framework, analyze reports, and create themes with models.}

\Revision{
\textbf{Select an attack technique framework.} 
We deductively code attack techniques using \acrlong{attck} (MITRE ATT\&CK)~\cite{mitreATTACK}.
We used the provided taxonomy of tactics and techniques used by adversaries across the attack life cycle. 
\Revision{Tactics} describe the goals (``why'') and techniques detail the actions to achieve the goal (``how'').}
In some cases, techniques detail specific descriptions of adversarial behavior in sub-techniques.
The identifiers are of the form \textit{TAXXXX} for tactics, \textit{TXXXX} for techniques, and \textit{TXXXX.YYY} for sub-techniques.
For example, attackers gain initial access (TA0001) to systems by leveraging trusted relationships (T1199).
\Revision{\gls{attck} enterprise version v16.1 has 14 tactics, 203 techniques, and 453 sub-techniques.}
We selected \gls{attck} because it:
(a) provides granular  \glsfirst{attack};
(b) is widely adopted by software security professionals; and,
(c) is used for threat intelligence.

\MandatoryRevision{\textbf{Analyze reports.}
We start our qualitative analysis by structuring our coding through a pilot.
We selected five reports (four from \gls{xz} and one from \gls{solarwinds}) that we considered of high quality, which two authors then qualitatively coded.
Based on our experience, we extract: adversarial events, non-adversarial events, and task recommendations.
Events, the basis of all incident analysis approaches, are defined as system state changes important enough to name~\cite{wienen2017accident}.
In our case, adversarial events are events conducted by attackers. 
We give details about non-adversarial events and task recommendations in Sec.~\ref{SEC:Metho-Map}.
Afterward, adversarial events are mapped to the 203 \gls{attck} attack techniques and 453 sub-techniques, using explicitly provided identifiers if given.
For example, an adversarial event is when \textit{``APT29 [adversaries] was able to get SUNBURST signed by SolarWinds code signing certificates by injecting the malware into the SolarWinds Orion software lifecycle''}, and mapped to subvert trust controls (T1553).}

\MandatoryRevision{
Our qualitative analysis procedure of the 106 reports was as follows.
First, two researchers independently qualitatively analyzed the 84 reports from strategies G1 and G2 (Sec.~\ref{SEC:Metho-Col}) following the structure and decided upon elements from the pilot.
Each author independently read each report, collected events, organized events in timelines, mapped adversarial events to \gls{attck} attack techniques, and revisited reports.
After independently analyzing the 84 reports, the two researchers met to unify the adversarial events and mappings to \gls{attck} attack techniques, negotiating disagreements~\cite{campbell2013coding}.
We seek agreement rather than inter-rater reliability, as disagreements allowed us to improve points of confusion or tension~\cite{mcdonald2019reliability}.
Most of our disagreements arose from the granularity of the events and ambiguities in the reports about the \gls{aus}.
When necessary, the authors consulted references in the reports and discussed issues with the team.
As we strive for saturation, in the next phase, two researchers analyzed the remaining 22 reports from the G3 strategy (Sec.~\ref{SEC:Metho-Col}).
We analyzed the remaining reports using the same procedure, and we reached theoretical saturation from the attacks.
Finally, the first author revisited the attack technique mappings to improve consistency, which the second author checked.
We found 316 unique adversarial events in \gls{solarwinds}, 45 in \gls{log4j}, and 168 in \gls{xz} from the 106 reports.
} 

\begin{figure*}[tb]
    \centering
      \begin{subfigure}[b]{0.40\linewidth}
         \centering
        \includegraphics[width=\linewidth]{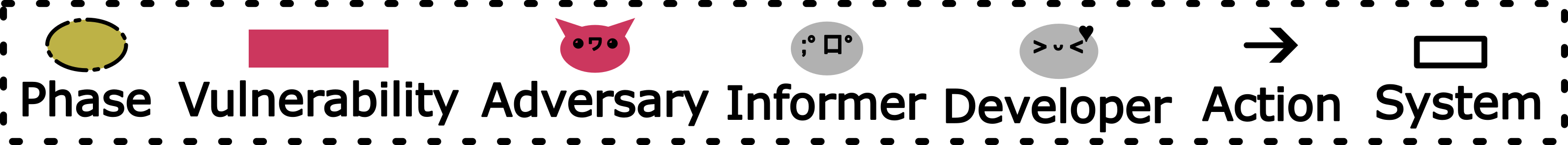}
     \end{subfigure}\\
     \begin{subfigure}[b]{0.33\textwidth}
         \centering
        \includegraphics[width=\linewidth]{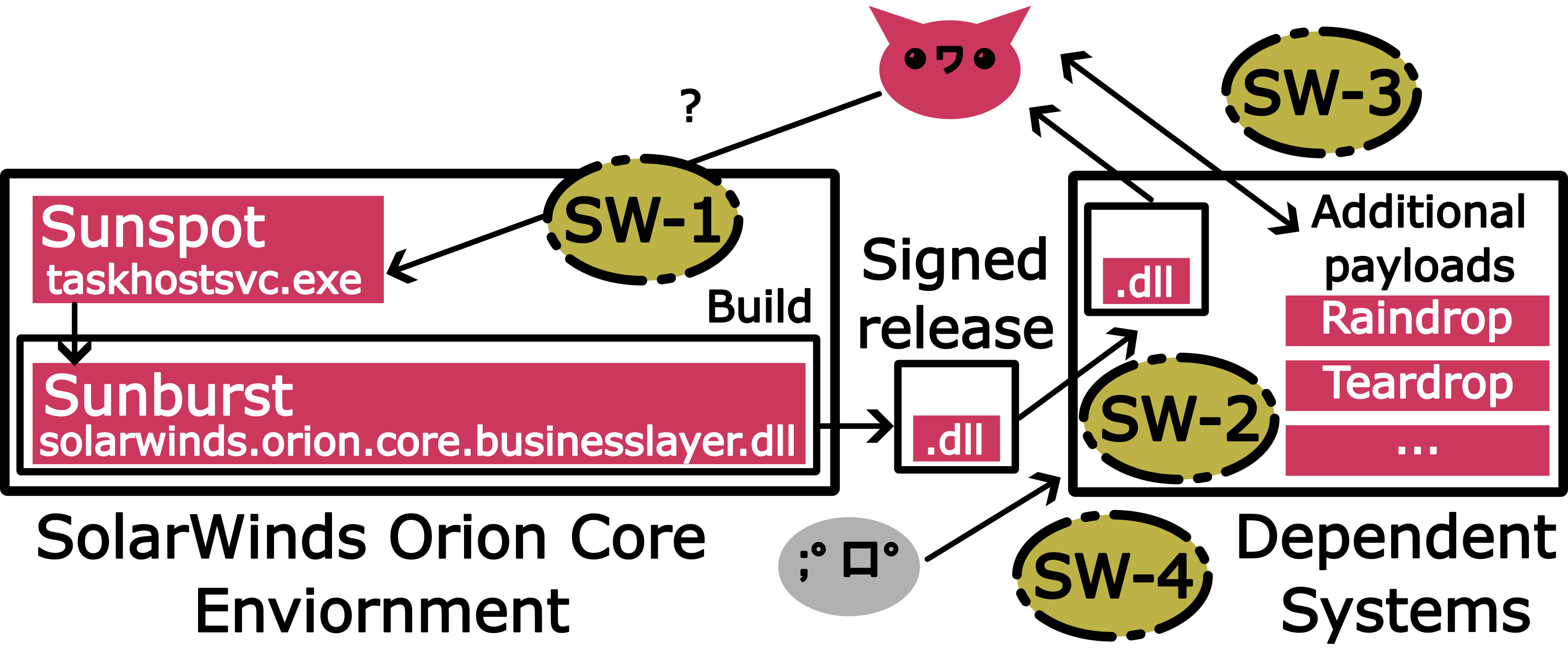}
         \caption{\gls{solarwinds}}\label{fig:tm-sw}
         \Description[SolarWinds Threat Model]{}
     \end{subfigure}
     \hfill
     \begin{subfigure}[b]{0.33\textwidth}
         \centering
         \includegraphics[width=\linewidth]{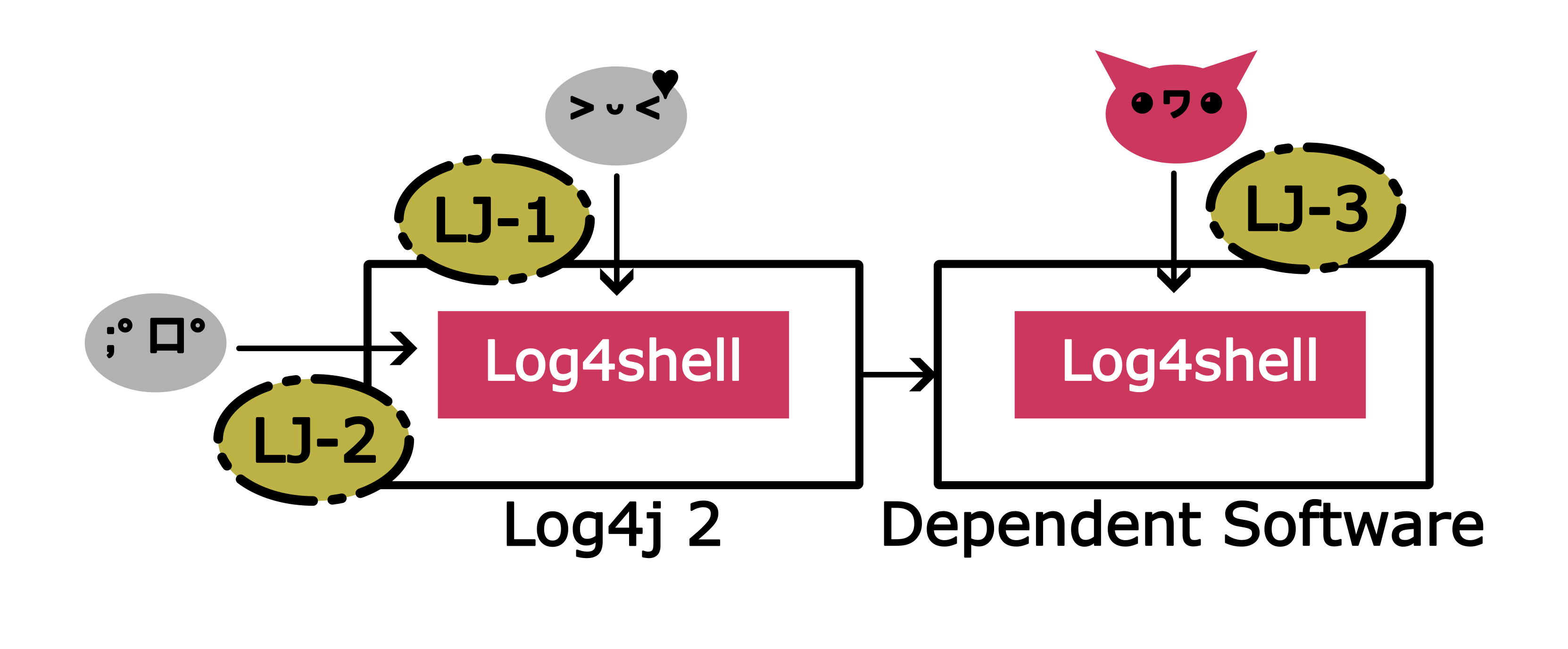}
         \caption{\gls{log4j}}\label{fig:tm-log4j}
         \Description[Log4j Threat Model]{}
     \end{subfigure}
     \hfill
     \begin{subfigure}[b]{0.33\linewidth}
         \centering
     \includegraphics[width=\linewidth]{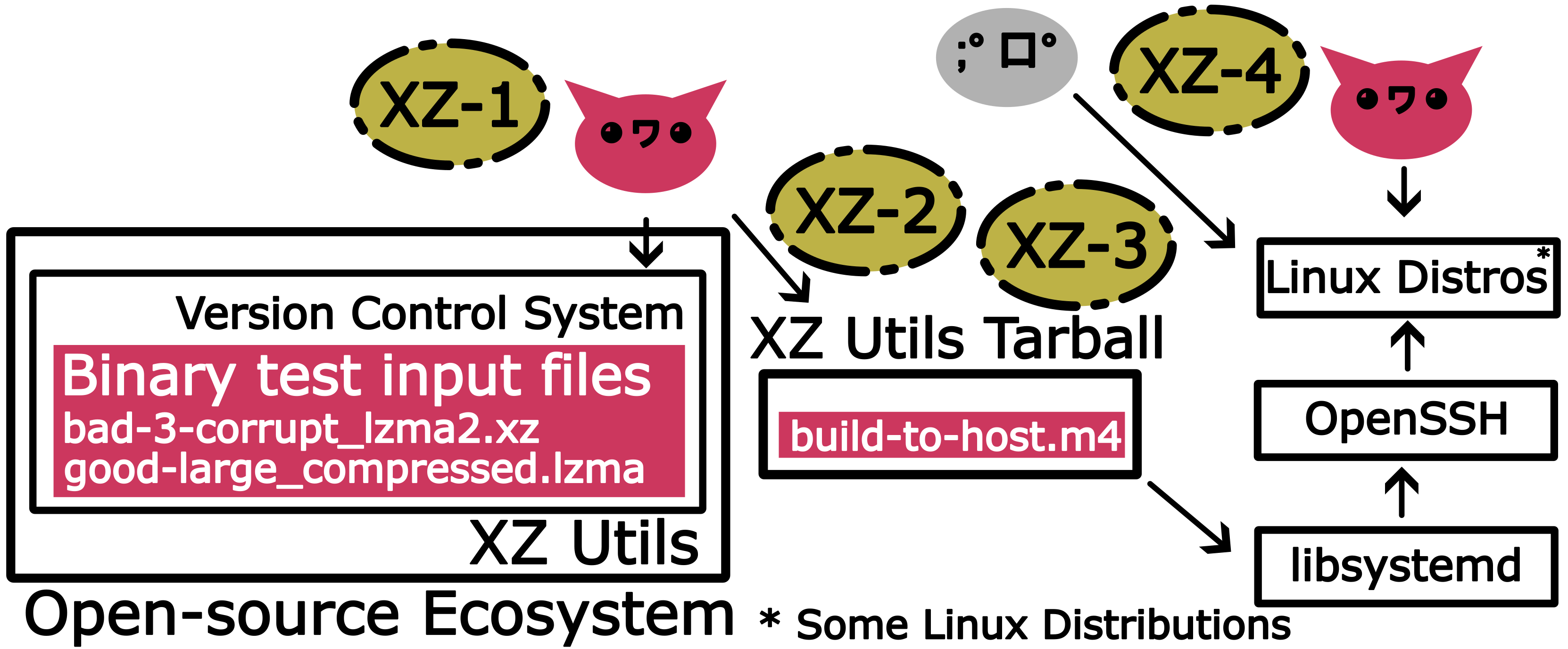}
         \caption{\gls{xz}}\label{fig:tm-xz}
          \Description[XZ Utils Threat Model]{}
     \end{subfigure}

    \caption{\Revision{Threat models of the software supply chain attacks}}\label{fig:threat-model}
\end{figure*}

\MandatoryRevision{\textbf{Create themes with models.}
After, we grouped events into themes, which we call stages, to provide a broader view of the \gls{aus}. 
We additionally created threat models of the stages shown in Fig.~\ref{fig:threat-model} and discussed in Sec.~\ref{SEC:RQ1}.
Threat models represent high-level threats to systems~\cite{threatModeling}.
We additionally tagged at what level the adversarial event compromised the software supply chain as: dependency, dependent, or both.
For example, we classified the prior example adversarial event to the \textit{Sunburst} introduction phase and tagged the compromise at the dependency level.
The authors discussed and created the themes.
The first author then annotated the themes, and the second author checked all annotations.
}

\subsection{Map Attack Techniques to Tasks}\label{SEC:Metho-Map}

\MandatoryRevision{
We map attack techniques with framework tasks to determine how to prevent future incidents~\cite{leveson2016engineering}.
\MandatoryRevision{As we want to improve the security posture of software organizations and frameworks, we map mitigation tasks based on the latest frameworks.} 
We curate an attack technique to task mapping through three steps: select a task framework, collect attack techniques to task mappings, and select attack techniques to task mappings.}

\begin{table}[t]
\centering
\caption{Ten referenced frameworks and \textit{\gls{psscrm}}}\label{TAB:PSSCRM}
\begin{tabular}{@{}llrrrrr@{}}
\toprule
&  & \multicolumn{4}{c}{\textbf{Tasks in group}} \\
\textbf{Framework}               & \textbf{Developed} & \textbf{G} & \textbf{P} & \textbf{E} & \textbf{D} \\ \midrule
\acrshort{nist} 800-161~\cite{800161}                           & Government & \cellcolor{myblue!30!white} 20 & 10 & 9 & 5 \\
\acrshort{eo} 14028~\cite{eo14028}/\acrshort{ssdf}~\cite{ssdf}  & Government & 12 & \cellcolor{myblue!30!white} 17 & 6 & 7 \\
\acrshort{attestation}~\cite{selfAttestation}                   & Government & 9 & \cellcolor{myblue!30!white} 11 & 5 & 4 \\
\textit{\gls{psscrm}}~\cite{williams2024proactive}              & Academia & \cellcolor{myblue!30!white} 23 &  19 & \cellcolor{myblue!30!white} 23 & 8 \\
\acrshort{scorecard}~\cite{scorecardOpenSSF}                    & Open-source & 0 & \cellcolor{myblue!30!white} 6 & 2 & 1 \\ 
\acrshort{bsimm}~\cite{bsimm}                                   & Industry  & \cellcolor{myblue!30!white} 17 & 14 & 2 & 4 \\
\acrshort{cncfSSC}~\cite{cncfSSCF}                              & Open-source & 4 & 6 & \cellcolor{myblue!30!white} 13 & 1 \\
\acrshort{owaspSCVS}~\cite{owaspSCVS}                           & Open-source & 1 & \cellcolor{myblue!30!white} 5 & \cellcolor{myblue!30!white} 5 & 0 \\
\acrshort{s2c2f}~\cite{s2c2f}                                   & Open-source & 3 & \cellcolor{myblue!30!white} 7 & 3 & 2 \\
\acrshort{slsa}~\cite{slsa}                                     & Industry & 2 & 1 & \cellcolor{myblue!30!white} 3 & 0 \\
\bottomrule
\multicolumn{6}{c}{\footnotesize \textit{Group:} P = Product , E = Environment, G = Governance, D = Deployment}\\
\multicolumn{6}{c}{\footnotesize \colorbox{myblue!30!white}{Framework focus indicated by the group with the highest number of tasks}}
\end{tabular}
\end{table}

\Revision{
\textbf{Select a task framework:} 
We leverage the  Proactive Software Supply Chain Risk Management (P-SSCRM)~\cite{williams2024proactive} framework to deductively code software supply chain tasks.}
The framework maps bi-directionally equivalent tasks described in ten prominent referenced risk management software supply chain frameworks shown in Table~\ref{TAB:PSSCRM}.
Two tasks are bi-directionally equivalent~\cite{mappingNIST} if \Revision{the tasks} have the same meaning but use different wording or phrasing in the framework definitions. 
Each framework in \gls{psscrm} has a focus, as indicated by the group column, with the union of all providing a holistic view of software supply chain security.
Note that \acrshort{ssdf}~\cite{ssdf} contains the \gls{eo} 14028~\cite{eo14028} tasks.
Each \gls{psscrm} task belongs to a group (high-level objective) and a practice (mid-level objective). 
\Revision{The latest \gls{psscrm} version v1.01 contains 4 groups, 15 practices, and 73 tasks.}
\Revision{In the task identifier (X.Y.Z), the X indicates the group, Y the practice, and Z the task.}
For example, to reduce risk in the product (P), vulnerable components and containers can be managed (P.5) by consuming \glspl{sbom} (P.5.1).
We selected \gls{psscrm} because it:
(a) contains only tasks for software supply chain attacks; and
(b) maps to 10 prominent reference frameworks in unified tasks.

\Revision{
\textbf{Collect attack technique to task mappings.} 
\gls{psscrm} does not directly map to the attack techniques mitigated.
We therefore created a \gls{attck} attack technique to \gls{psscrm} task mapping.}
Creating a manual mapping between \gls{attck} attack techniques and \gls{psscrm} tasks is complex as there are, without the gaps found (Sec.~\ref{SEC:Metho-Gap}), $14,819$ possible combinations (203 attack techniques $\times$ 73 tasks).
Additionally, mapping \Revision{cybersecurity} frameworks is subjective~\cite{mappingNIST},
thus, triangulation is needed to find where opinions converge.
\MandatoryRevision{Triangulation is a technique that strengthens research design by using different research approaches~\cite{thurmond2001point}.
In our case, we use methodological triangulation using four independent, systematic, and scalable strategies, which are: transitive (M1), \gls{llm} (M2), framework (M3), and reports (M4).
We describe each strategy in the following paragraphs.}

\textbf{\textit{Transitive mappings (M1):}} 
\Revision{Although there is no direct \gls{attck} attack technique to \gls{psscrm} task mapping, transitive mappings between \Revision{frameworks can be found}.
For example, we can \Revision{map} \gls{attck} attack techniques $\leftrightarrow$ \gls{nist} SP 800 53  $\leftrightarrow$ \gls{nist} \gls{ssdf} $\leftrightarrow$ \gls{psscrm} tasks.}
\Revision{Hence, we constructed a dataset to leverage the transitive mappings}.
\Revision{We start by finding the mappings to \gls{attck} and the ten referenced frameworks.}
We iteratively searched through web pages, Google searches, and documentation for mappings from and to the frameworks and found 183 framework-to-framework mappings.

We used the \textit{networkx} package in \textit{Python} to find all simple paths~\cite{networkx} between \gls{attck} and \gls{psscrm} to find paths with no repeated frameworks.
We set a cutoff value \Revision{for} the path length as the number of simple paths increases exponentially.
We discarded superset simple paths as we wanted the shortest simple paths. 
For example, we discard the path \gls{attck} $\leftrightarrow$ \gls{nist} SP 800 53  $\leftrightarrow$ \gls{nist} \gls{ssdf}  $\leftrightarrow$ \gls{nist} 800 161 $\leftrightarrow$ \gls{psscrm} as the simple path was a superset of our prior example.
As more than $99\%$ of simple paths with a length above $7$ superset another shorter path, we selected $10$ as a conservative threshold.
We remained with 33 frameworks and 78 framework-to-framework mappings that generated 125 simple paths.
We then collected the data, cleaned up inconsistencies, and found 17 simple paths with data.
For the 108 simple paths without data, we manually checked the paths and found no data to create a transitive path in most cases (101 paths).
For example, although a simple path exists between \gls{attck} $\leftrightarrow$ OWASP Community Attacks $\leftrightarrow$ OWASP Cheat Sheets, no attack technique maps to an OWASP cheat sheet.
Other reasons were differences in the data granularity (5 paths) and no end-to-end transitive data (2 paths).
We provide the framework-to-framework mappings, simple paths, and data in our supplemental material~\cite{supplementalLink}.
We found 2,534 candidate attack technique to task mappings.

\textbf{\textit{\gls{llm} mappings (M2):}} 
We leveraged \gls{llm} capabilities to find patterns in data to create attack technique to task mappings.
We tested three models: ChatGPT gpt-4o-mini~\cite{chatgpt}, VertexAI with gemini-1.5-pro~\cite{vertexAI}, and Claude 3.5 Sonnet 2024-10-22~\cite{claude}.
We selected the models as they provided APIs used in prior work~\cite{zahan2024shifting, bae2024enhancing} or included grounded models~\cite{grounded}.
To reduce bias in our prompt, available in supplemental material~\cite{supplementalLink}, we base the wording using the US NIST's \Revision{cybersecurity} mapping standard~\cite{mappingNIST} and the frameworks we are mapping~\cite{mitreATTACK,williams2024proactive} using a \gls{cot} prompt inspired from prior work~\cite{dunlap2024pairing}.
We ask to find bi-directionally supportive relationships where a task mitigates an attack technique.
We also ask the prompt to produce a binary answer~\cite{tai2024examination} for a pair of attack technique to task mapping at a time to increase accuracy.

We choose the \gls{llm} based on a sample of attack technique to task mappings.  
We stratified our sample~\cite{baltes2022sampling} based on the combinations of \gls{attck} tactics and \gls{psscrm} groups to ensure we get pairs of each group.
We then randomly selected, proportional to the size of the number of pairings, 150 samples to have at least one pair for each strata.
For example, without the gaps (Sec.~\ref{SEC:Metho-Gap}), there are 44 attack techniques in the defense evasion (TA0005) tactic and 23 tasks in the governance (G) group leading to $44 \times 23 = 1,012$ possible pairings.
Considering some attack techniques belong to multiple tactics, the number of possible pairings across all groups is $73 \times  236 =17,228$.
The number of pairs for the strata by proportional sampling is $\lceil\frac{1,012}{17,228} \times 150 \rceil = 9$.
We tested both zero-shot and one-shot versions of the prompt in each model, generating 900 \gls{llm} responses.
Two authors independently mapped if the task mitigated the  attack technique for the 150 samples and negotiated disagreements. 
We selected ChatGPT GPT-4o-mini as the model with a zero-shot prompt as: 
(a) zero-shot outperformed one-shot, 
(b) 82\% of the pairs were the same as the disagreement-resolved sample, and
(c) the cost was only 3\% of the other models.
We, therefore, generated 2,259 attack technique to task mappings.

\textbf{\textit{Framework mappings (M3):}} 
\Revision{The \gls{attck} framework also contains general software security mitigations, which we call tasks in this paper, that map to attack techniques.}
Thus, we map \gls{attck} and \gls{psscrm} tasks and leverage the existing mapping (\gls{attck} attack techniques $\leftrightarrow$ \gls{attck} tasks $\leftrightarrow$ \gls{psscrm} tasks).
We followed the NIST mapping standard~\cite{mappingNIST}, mapping set relationships between \gls{attck} and \gls{psscrm} tasks\Revision{,} as there were few bi-directionally equivalent tasks.
The first author created the  \gls{attck} and \gls{psscrm} task set mapping, which was then reviewed by the second author.
\Revision{We collected 548 candidate attack technique to task mappings.}

\textbf{\textit{Report mappings (M4):}} 
From the qualitative analysis (Sec.~\ref{SEC:Metho-Qual}), we also extracted non-adversarial events and task recommendations manually mapped to the mitigated attack technique.
For example, the non-adversarial event \textit{``as part of our [\gls{solarwinds}] response to the SUNBURST vulnerability, the code-signing certificate used by SolarWinds to sign the affected software versions was revoked...''} is mapped to the subvert trust controls (T1533) and trusted relationship (T1199) attack techniques with the software release integrity (P.1.2) task.
When available, we also extracted recommended tasks, which we defined as suggestions for mitigating the attacks.
For example, a task recommendation in \gls{solarwinds} is \textit{``organizations need to harden their build environments against attackers''} is mapped to the subvert trust controls (T1533) attack technique with CI/CD hosting and automation (E.2.4) task.
We found 14 unique non-adversarial events in \gls{solarwinds}, 58 in \gls{log4j}, and 53 in \gls{xz}.
Meanwhile, we found 653 task recommendations in \gls{solarwinds}, 791 in \gls{log4j}, and 180 in \gls{xz}.
After removing duplicates, we collected 296 candidate attack technique to task mappings.

\textbf{Select attack technique to task mappings}.
We found 5,628 candidate attack technique to task mappings from the four strategies, which resulted in 4,366 unique mappings after removing duplicates.
\MandatoryRevision{We calculated Krippendorff's alpha to determine the agreement between strategies, resulting in a value of 0.1, indicating slight agreement~\cite{gonzalez2023reliability}.
Hence, strategies are not substituting for each other, aligning with methodological triangulation~\cite{thurmond2001point}. 
We thus filter the attack technique to task mappings to select only agreed-upon mappings.
}
We selected three strategies as our threshold to find where opinions converged while allowing us to find some differences between strategies.
We, therefore, selected 235 attack technique to task mappings available in our supplemental material~\cite{supplementalLink}.

\subsection{\MandatoryRevision{Map Reports to Missing Tasks}}\label{SEC:Metho-Gap}

\MandatoryRevision{
We gathered framework gaps by qualitatively analyzing the 106 reports (Sec.\ref{SEC:Metho-Col}).
We intended to identify tasks missing from the ten referenced frameworks to consolidate the tasks into the software supply chain frameworks that software organizations adopt.
First, the two authors created candidate framework gaps when we found no appropriate \gls{attck} attack techniques (Sec.~\ref{SEC:Metho-Qual}) and \gls{psscrm} tasks (Sec.~\ref{SEC:Metho-Map}) when mapping the reports.}
Afterward, six authors met throughout several meetings to discuss: 
(a) if the additions were actual framework gaps;
(b) how the changes would fit into the frameworks; and
(c) the magnitude of the changes.
We classified the magnitude of the changes into three types:
(a) \textit{full}, if a new attack technique or task was created; 
(b) \textit{major}, if an existing attack technique or task scope was expanded; 
and
(c) \textit{minor}, if only the writing of the attack technique or task was changed.
\Revision{The disagreements~\cite{mcdonald2019reliability} arose from the diversity of perspectives from the authors and the scope of tasks in \gls{psscrm}.}
The first author compiled the discussions into changes for \gls{attck} and \gls{psscrm}, which the authors then discussed and refined.
\Revision{The first author then re-mapped the gaps in the reports, with another author reviewing the changes.}

We performed two validity checks for our gaps.
First, to ensure that the gaps were due to omissions in the ten referenced frameworks rather than mistakes in \gls{psscrm}, two authors checked the frameworks: 
(a) if there is no bi-directionally equivalent task for the \textit{full} gaps; 
(b) if the expanded scope existed in the task for the \textit{major} gaps; and
(c) if the changed wording was absent in the referenced task for the \textit{minor} gaps.
The two authors then discussed each gap while consulting other authors and discarded gaps due to mistakes in \gls{psscrm}.
Second, we contacted the authors of \gls{attck} and \gls{psscrm} for feedback on the gaps.
\Revision{Both framework authors responded that the gaps are in queue for a future release.}
\Revision{The suggested changes are available in the supplemental material~\cite{supplementalLink}.}

Additionally, we constructed the attack technique to task mappings (Sec.~\ref{SEC:Metho-Map}) for the full gaps using the M2, M3, and M4 strategies.
We could not apply the transitive strategy (M1) as none of the ten referenced frameworks mapped to the full gaps.
\Revision{Thus, we chose the convergence to be two to allow some strategy differences.}
We selected 16 attack techniques to full gaps mappings shared in the supplemental material~\cite{supplementalLink}.

\subsection{\Revision{Map AUS to Mitigation Tasks}}\label{SEC:Metho-Val}

To establish the mitigation tasks, we combined the attack techniques of each \gls{aus} (Sec.~\ref{SEC:Metho-Qual}) with the attack technique to task mappings (Sec.~\ref{SEC:Metho-Map} and~\ref{SEC:Metho-Gap}). 
\Revision{We thus create a transitive mapping as \gls{aus} attack techniques $\leftrightarrow$ \gls{attck} attack techniques $\leftrightarrow$ \gls{psscrm} tasks and gap tasks}.
\Revision{Aligned with software security rankings such as OWASP Top Ten~\cite{owasptop}, we rank tasks based on the current trends of the \gls{aus}.}
For each mitigation task, we calculate a score to rank the tasks by multiplying the number of \gls{aus} mitigated by the task with the number of attack techniques mitigated by the task.
For example, the score of multi-factor authentication (E.1.3) is four as the task mitigates two \gls{aus} (\gls{solarwinds}, \gls{log4j}) and two attack techniques (T1078, T1098).
\Revision{Finally, we define our \gls{starter} as the essential tasks a software organization should prioritize adopting to mitigate attack techniques.}
We selected the mitigation task with the top ten scores, as shown in Table~\ref{tab:priority}.
Ten was selected in line with other software security rankings like the OWASP Top Ten~\cite{owasptop}.

\subsection{Measure Frameworks}\label{SEC:Metho-Measure}

\MandatoryRevision{We collect three measures for each software supply chain framework: hit-count, coverage, and completeness.
using the attack technique to task mappings (Sec.~\ref{SEC:Metho-Map}), we calculate \textit{hit-count} as the number of framework tasks that mitigate attack techniques.
For example, \gls{nist} 800-161 tasks mitigate 91 attack techniques out of 203 in \gls{attck}. 
To calculate framework \textit{coverage} and \textit{completeness}, we first}
classify each mitigation task (Sec.~\ref{SEC:Metho-Val}) as matched, unused, or missing for the ten reference frameworks and \gls{psscrm}.
A task is considered \textit{matched} if it is: (a) included in the framework, and (b) mitigates an attack technique for the \gls{aus}.
For example, \gls{nist} 800-161 includes role-based access control (E.3.3), which mitigates attack techniques in \gls{solarwinds}. 
A task is considered \textit{unused} if it is: (a) included in the framework, and (b) does not mitigate an attack technique for the \gls{aus}.
For example, \gls{nist} 800-161 includes the software license conflict (G.1.2) task, yet the task does not mitigate attack techniques in \gls{solarwinds}.
Finally, a task is considered \textit{missing} if it is: (a) not included in the framework, and (b) mitigates an attack technique for the \gls{aus}.
For example, \gls{nist} 800-161 does not include the secure design review (P.2.1) task that mitigates attack techniques in \gls{solarwinds}.
\Revision{Finally, inspired by precision and recall, we calculate for each framework:}
(a) \textit{coverage} -- the number of matched tasks divided by the sum of matched and unused tasks; and
(b) \textit{completeness} -- the number of matched tasks divided by the sum of matched and missing tasks.

\section{\rqref{rq:events} AUS Attack Techniques}\label{SEC:RQ1}

\textbf{\A{rq:events-text}} We answer RQ1 by analyzing the \gls{attck} attack techniques and tactics found from our qualitative analysis presented in Sec.~\ref{SEC:Metho-Qual}.
We narrate the phases with events for the \gls{aus} and present common trends across all \gls{aus}.
The \gls{aus} threat models are shown in Fig.~\ref{fig:threat-model} and the full list of attack techniques is in the supplemental material~\cite{supplementalLink}.
New \gls{attck} attack techniques are identified using T-newX, where X is a number.

\begin{figure*}[tb]
    \centering
     \centering
     \begin{subfigure}[b]{0.24\textwidth}
         \centering
        \includegraphics[width=\linewidth]{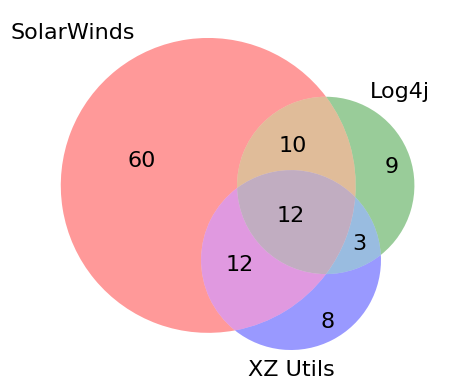}
         \caption{Overlap of attack techniques}
         \label{fig:technique-overlap}
          \Description[Overlap of attack techniques]{}
     \end{subfigure}
     \hfill
     \begin{subfigure}[b]{0.22\textwidth}
         \centering
         \includegraphics[width=\linewidth]{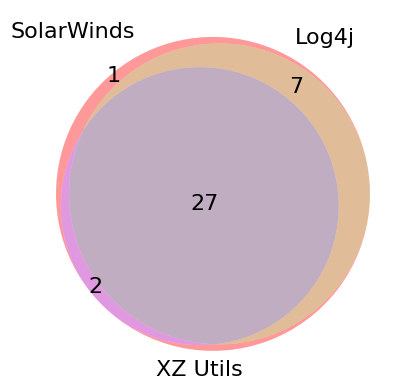}
         \caption{Overlap of mitigation tasks}
         \label{fig:task-overlap}
         \Description[Overlap of mitigation tasks]{}
     \end{subfigure}
     \hfill
     \begin{subfigure}[b]{0.53\linewidth}
         \centering
      \includegraphics[width=.9\linewidth]{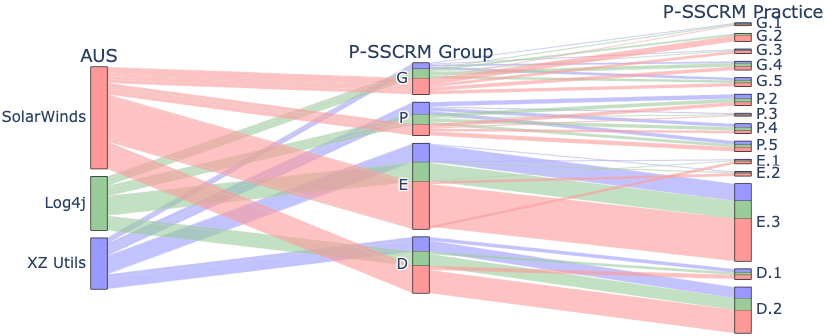}\\
      {\footnotesize *Size is the number of \gls{attck} attack techniques related to mitigation tasks}
         \caption{Relationship between \gls{aus}, \gls{psscrm} groups, and practices}
         \label{fig:relationship}
        \Description[Relationship between \gls{aus}, \gls{psscrm} groups, and practices]{}
     \end{subfigure}
    \caption{Trends in the \gls{attck} attack techniques and \gls{psscrm} tasks for \gls{solarwinds}, \gls{log4j}, and \gls{xz}.}
    \label{fig:trends}
\end{figure*}

\textbf{\gls{solarwinds}.} 
We found four stages in the incident (Fig.~\ref{fig:tm-sw}): \textit{Sunspot} introduction (SW-1), \textit{Sunburst} introduction (SW-2), dependent exploitation (SW-3), and vulnerability discovery (SW-4).
The adversaries leveraged 94 attack techniques, using all 14 tactics. 

The attack started in September 2019 when the adversaries accessed the \gls{solarwinds} Orion Core environment (SW-1).
The adversaries added the \textit{Sunspot} malware, a file named \textit{taskhostsvc.exe}, used to insert \textit{Sunburst}, a file named  \textit{solarwinds.orion.core.businesslayer.dll} into the software build (T1195).
The malware was likely developed by the adversaries (T1587).
Using debugging privileges granted by the malware (T1134), \textit{Sunspot} would monitor the \textit{MsBuild.exe} (T1057) process used by \textit{Visual Studio Code} when software is being built.
Afterward, \textit{Sunspot} would check that the backdoor code is compatible with a known source code file using an \textit{MD5 checksum} (T1480).
If so, the backdoor would then decrypt the \textit{AES128-CBC} encrypted \textit{Sunburst} (T1140) and modify the source code to contain the malware (T1565), taking advantage of the build tools (T1072).
\textit{Sunspot} masqueraded as a legitimate file with reasonable code (T1036), encrypted the logs generated (T1027), and removed created files (T1070).
The adversaries first tested if source code modifications would be detected by developers by inserting empty classes (T1622) and started inserting \textit{Sunburst} in February 2020. 
How \textit{Sunspot} was introduced in the environment is still unknown.

From February to June 2020, the trojanized versions of \gls{solarwinds} were available (SW-2) that included the staged malware \textit{Sunburst} (T1608) signed by \gls{solarwinds} (T1553).
Thousands of \gls{solarwinds} dependents in public and private sectors adopted the trojanized version (T1199), which then lay dormant for up to two weeks in the systems (T1497).
The malware would perform tasks the adversaries gave through command line arguments (T1059), including collecting system information (T1005), spawning processes (T1543), killing tasks (T1112), writing registry keys (T1112), and sending data (T1071, T1041).
The traffic sent was encrypted (T1573), encoded (T1132), contained junk bytes (T1001), obfuscated (T1027), and matched the victim's home country IP address (T1665, T1036).
The malware collected process names hashed against hardcoded blocklists (T1057) and attempted to disable security services (T1562).
\textit{Sunburst} would then deliver additional payloads (T1105).

The adversaries started exploiting the dependents through the additional payloads (SW-3).
Example payloads included \textit{Teardrop}, \textit{Raindrop}, \textit{Cobalt Strike}, and \textit{Sunshuttle}.
The additional payloads used similar attack techniques as \textit{Sunspot} and \textit{Sunburst}, gathering information about the systems then exfiltrated while evading detection mechanisms.
However, further attack techniques were used to exploit the systems.
For example, credentials were stolen (T1078, T1539, T1558, T1589), manipulated (T1098, T1134, T1558), forged (T1550, T1606, T1558), compromised from services (T1555), and abused (T1548). 
The adversary accessed internal repositories (T1213), copied data (T1003), captured screens (T1113), collected emails (T1114), and extracted files (T1005).
In December 2020, FireEye, a cyber security company,
discovered the attack (SW-4).

\textbf{\gls{log4j}.}
We found three stages in the incident (Fig.~\ref{fig:tm-log4j}): vulnerability introduction (LJ-1), vulnerability disclosure (LJ-2), and post-disclosure (LJ-3).
We found 34 attack techniques and all 14 tactics.

The vulnerability was introduced (LJ-1) in 2013 when a member of the open-source community requested to add a \gls{jndi} lookup feature to ease data storage and retrieval of remote sources not part of the installed application.
The same community member developed the feature in a patch accidentally containing vulnerable code.
The \gls{log4j} development team reviewed and committed the patch, scheduling the feature for version 2.0-beta9.6.
In November 2021, Chen Zhaojun, a security developer at Alibaba, reported the vulnerability in the \gls{jndi} feature to Apache (LJ-2). 
The Apache team started to devise a fix, tagging as a release candidate Log4j-2.15.0-rc1.
However, adversaries were discussing in open forums (T1593) \gls{poc} of the exploit (T1588).
Chen Zhaojun informed Apache the vulnerability was being discussed in forums. 
Thus, Apache publicly disclosed the vulnerability in CVE-2021-44228 and released a fix ahead of the scheduled version 2.15.0 in December 2021.

Subsequently, adversaries started exploiting the vulnerability (LJ-3).
Adversaries gained initial access to systems through the \gls{log4j} vulnerability (T1195), which was as a widely used logging library by Apache and thus subsequently trusted (T1199).
The adversaries used specially crafted input to inject content (T1659) to exploit the vulnerability.
Several additional vulnerabilities were found and subsequently exploited.
An incomplete fix for CVE-2021-44228 resulted in CVE-2021-45046 being discovered, while CVE-2021-45105 and CVE-2021-44832 were discovered for other vulnerabilities on top of CVE-2021-45046.
Additionally, CVE-2021-4104 was discovered, affecting the deprecated Log4j version 1.
Applications were scanned to identify vulnerable systems (T1595), which resulted in the exploitation of public-facing applications (T1190).
Adversaries used the vulnerabilities to enable attacks, including remote code execution (T1210), arbitrary code execution (T1203), data leakage (T1041), botnet malware (T1584), cryptocurrency miners (T1496), denial of service (T1499), and ransomware payloads (T1486).
Other existing vulnerabilities were also leveraged in attacks (T1588) and additional payloads uploaded (T1105).
The adversaries evaded defense mechanisms (e.g., T1027, T1036, T1078) and established persistence mechanisms (T1053, T1505, T1078) within the attacked systems.

\textbf{\gls{xz}.}
We grouped the incident into four stages (Fig.~\ref{fig:tm-xz}): became maintainer (XZ-1), backdoor introduction (XZ-2), backdoor improvement (XZ-3), and vulnerability discovery (XZ-4).
We found 35 \glsfirst{attack}, using 11 tactics. 
The excluded tactics were collection (TA0009), exfiltration (TA0010), and impact (TA0040).

In October 2021, the adversaries started to become maintainers of \gls{xz} (XZ-1).
Jia Tan, an identity used by the adversaries, created a sock puppet account on GitHub (T1585), possibly impersonating someone (T1656).
The developer started by cultivating relationships (T1585) in the \gls{xz} and \textit{libarchive} projects by contributing patches.
Jia Tan also tested \textit{libarchive} capabilities (T1622) by introducing a vulnerability by replacing a safe print with an unsafe variant.
In April 2022, Jia Tan first sent another patch to \gls{xz} mailing list. 
Then, Jigar Kumar and Deniss Ens, suspected sock puppet accounts (T1585), sent emails for several months pressuring (T-new1) Lasse Collin, the maintainer of \gls{xz}, to make Jia Tan a maintainer.
By June 2022, Lasse Collin informs that maintainer changes are in progress.
Jia Tan was then added to the GitHub organization (T1072) and continued contributing (T1585), becoming deeply involved and aware of the security of \gls{xz} (T1592).
Additionally, Jia Tan changed the primary email address in Google's \textit{oss-fuzz} of \gls{xz} to themselves (T1548).

The adversaries then started contributing changes related to the backdoor (XZ-2).
The complete exploit was introduced in February 2024 when Jia Tan took advantage of binary test files that were stored in a repository directory (T1553) and added two binary test input files for the backdoor (T1027): \textit{bad-3-corrupt\_lzma2.xz} and \textit{good-large\_compressed.lzma}.
Jia Tan then published in v5.6.0 the complete backdoor (T1195), including a \textit{build-to-host.m4} file that activates the backdoor in the binary test input files (T1608).
The adversaries took advantage that \gls{xz} is a valuable compression utility dependency as several high-profile systems, such as Debian, are dependents (T1199).
Note that other changes were committed to the repository starting June 2023 to prepare for the backdoor (T1622).
The backdoor, in a series of rounds, ingress (T1105) obfuscated binary files (T1027) that are then deobfuscated (T1140) by string substitutions and offsets through shell scripts (T1059). 
The script then added lines to the \textit{Makefile} of \textit{liblzma} (T1608) to call a \textit{crc64\_resolve} function in \textit{liblzma\_la-crc64-fast.o} before trust controls for critical functions are set as read-only (T1553).
Thus, the values for \textit{RSA\_public\_decrypt} for sshd are changed to a nefarious version (T1543).
Hence, allowing adversaries to execute remote code (T1210) and allow interactive sessions (T1556) using specific (T1546) encrypted and signed payloads (T1573).
To evade detection, the backdoor is only activated on specific conditions (T1480), masqueraded as legitimate test files (T1036), disabled linker optimizations (T1562), and removed indicators (T1070).
Lasse Collin was on an internet break when the backdoor was introduced (T-new2).
Still, the backdoor had implementation issues that led to reported crashes in Linux distributions such as Red Hat and Gentoo.

A new version of the backdoor was developed (XZ-3) in v5.6.1 (T1608) in March 2024, fixing some of the \textit{ifunc} bugs, serving as a misdirection (T1036).
The backdoor was slightly improved, and among the changes, an extension mechanism (T1505) was added, allowing adversaries to add new test files and not regenerate existing ones for the attacks.
The files are found (T1083) offsetting bytes matched to a format.
New accounts were created (T1585) to pressure dependents to update to the new version (T-new1).
Still, due to the backdoor performance, Andres Freund, a Microsoft developer, notified and published a report about the attack in March 2024 (XZ-4) before exploitation of systems occurred.

\textbf{Across \gls{aus}.} 
We found 114 unique attack techniques out of the 203 \gls{attck} attack techniques in the \gls{aus}.  %
Out of the 114 attack techniques, 12 were leveraged in all 3 \gls{aus} as shown in Fig.~\ref{fig:technique-overlap}.
\Revision{We also gathered if the compromised component was a dependent, dependency, or both, for each adversarial event.}
The 12 attack techniques leveraged in only dependent components are: encrypted channels (T1573), remote services (T1021), ingress tool transfer (T1105), and command and scripting interpreter (T1059).
Meanwhile attack techniques leveraged in both components are: trusted relationships (T1199),  supply chain compromise (T1195), abuse elevation control mechanisms (T1548), obfuscated files or information (T1027), deobfuscation (T1140), masquerading (T1036), compromise infrastructure (T1584), and develop capabilities (T1587).

\begin{tcolorbox}[myhighlightbox]
    \Revision{
    \textbf{Key Insights:} Attack techniques used in software supply chain attacks are diverse (114 unique attack techniques across all AUS) and complex (a minimum of 34 attack techniques in an AUS).
    Still, 12 attack techniques also overlap across the AUS.}
\end{tcolorbox}

\begin{table}[t]
\centering
\caption{\Revision{Top ten mitigation \glspl{safeguard} in the \gls{starter}.}}\label{tab:priority}
\begin{tabularx}{.79\linewidth}{rcccccX}
\toprule
\textbf{Score} & \textbf{\Gls{safeguard}} & \textbf{SW} & \textbf{L4} & \textbf{XZ} & \textbf{Level} & \textbf{Gap}\\
\midrule
90 & E.3.3 & $\checkmark$ & $\checkmark$ & $\checkmark$ & $\updownarrow$ & $\CIRCLE$ \\
87 & D.2.1 & $\checkmark$ & $\checkmark$ & $\checkmark$ & $\updownarrow$ & $\RIGHTcircle$ \\
75 & E.3.7 & $\checkmark$ & $\checkmark$ & $\checkmark$ & $\updownarrow$ & $\RIGHTcircle$ \\
33 & E.3.6 & $\checkmark$ & $\checkmark$ & $\checkmark$ & $\updownarrow$ & $\CIRCLE$ \\
27 & E.3.11 & $\checkmark$ & $\checkmark$ & $\checkmark$ & $\updownarrow$ & $\Circle$ \\
18 & P.5.2 & $\checkmark$ & $\checkmark$ & $\checkmark$ & $\updownarrow$ & $\LEFTcircle$ \\
18 & P.2.1 & $\checkmark$ & $\checkmark$ & $\checkmark$ & $\updownarrow$ & $\CIRCLE$ \\
15 & E.3.4 & $\checkmark$ & $\checkmark$ & $\checkmark$ & $\updownarrow$ & $\CIRCLE$ \\
12 & G.2.6 & $\checkmark$ & $\checkmark$ &  & $\updownarrow$ & $\CIRCLE$ \\
12 & D.1.2 & $\checkmark$ & $\checkmark$ & $\checkmark$ & $\updownarrow$ & $\RIGHTcircle$ \\
\bottomrule
\end{tabularx}

{\footnotesize SW = SolarWinds, L4 = Log4j, XZ = XZ Utils.}\\
{\footnotesize $\checkmark$ : Mitigates \gls{aus}. Mitigation level: $\uparrow$ = Dependency, $\downarrow$ = Dependent, $\updownarrow$ = Both.}\\
{\footnotesize Gap: $\Circle$ = Full, $\LEFTcircle$ = Major, $\RIGHTcircle$ = Minor, $\CIRCLE$ = None.}
\end{table}

\begin{table}[t]
    \centering
    \caption{\MandatoryRevision{The hit-count (HC), coverage \% (CV), and completeness \% (CM) of the frameworks for the \gls{aus}.}}
    \label{tab:Frameworks}
    \begin{tabular}{@{}lrrrrrrr@{}}
    \textbf{}           & \multicolumn{2}{c}{\textbf{SW}} & \multicolumn{2}{c}{\textbf{L4}} & \multicolumn{2}{c}{\textbf{XZ}} \\
    \textbf{Frameworks} & \textbf{HC} & \textbf{CV}           & \textbf{CM}           & \textbf{CV}         & \textbf{CM}        & \textbf{CV}       & \textbf{CM}       \\
    \cmidrule(lr){1-1}  \cmidrule(lr){2-2} \cmidrule(lr){3-4} \cmidrule(lr){5-6} \cmidrule(lr){7-8}
\acrshort{nist} 800-161 & 91 &
  \cellcolor{mygreen!55!white}57        &
  \cellcolor{mygreen!55!white}68        &
  \cellcolor{mygreen!55!white}55        &
  \cellcolor{mygreen!55!white}71       &
  \cellcolor{mygreen!55!white}45        &
  \cellcolor{mygreen!55!white}69        \\
\acrshort{eo} 14028/\acrshort{ssdf} & 22 &
    51      &
    \cellcolor{mygreen!55!white}51      & 
    49 & 
    \cellcolor{mygreen!55!white} 53     & 
    \cellcolor{mygreen!55!white} 46     & 
    \cellcolor{mygreen!55!white} 59  \\
\acrshort{attestation} & 22 & 
    \cellcolor{mygreen!55!white}55      & 
    43                                  & 
    \cellcolor{mygreen!55!white}55      & 
    47                                  & 
    \cellcolor{mygreen!55!white}48      & 
    48     \\
\textit{P-SSCRM} & 97 &
    \cellcolor{mypink!55!white} 42      & 
    \cellcolor{mygreen!55!white} 92    & 
    \cellcolor{mypink!55!white} 39      & 
    \cellcolor{mygreen!55!white} 91    & 
    \cellcolor{mypink!55!white} 32      & 
    \cellcolor{mygreen!55!white} 90 \\
\acrshort{scorecard} & 9 & 
    \cellcolor{mygreen!55!white} 67      & 
    \cellcolor{mypink!55!white} 16       & 
    \cellcolor{mygreen!55!white} 56 & 
    \cellcolor{mypink!55!white} 15        & 
    44        & 
    \cellcolor{mypink!55!white} 14       \\
\acrshort{bsimm} & 18 &
    \cellcolor{mypink!55!white} 47      & 
    46                                  & 
    44                                  & 
    47                                  & 
    39                                  & 
    48                                  \\
\acrshort{cncfSSC} & 49 &
    48      &
    35                                  & 
    \cellcolor{mypink!55!white} 44      & 
    35      & 
    \cellcolor{mypink!55!white}30      & 
    28      \\
\acrshort{owaspSCVS} & 6 &
    50        & 
    \cellcolor{mypink!55!white}19       & 
    50                                  & 
    \cellcolor{mypink!55!white}21       & 
    36       &
    \cellcolor{mypink!55!white}17   \\
\acrshort{s2c2f} & 10 &
    \cellcolor{mypink!55!white}41       & 
    \cellcolor{mypink!55!white}19       & 
    \cellcolor{mypink!55!white}41       &
    \cellcolor{mypink!55!white}21       & 
    35      & 
    21       \\
\acrshort{slsa} & 2 & 
     50       & 
    \cellcolor{mypink!55!white} 8       & 
     50       & 
    \cellcolor{mypink!55!white} 9       & 
    \cellcolor{mypink!55!white} 33       & 
    \cellcolor{mypink!55!white} 7\\   
\bottomrule
\multicolumn{7}{c}{\footnotesize 
\colorbox{mygreen!55!white}{Top three scoring framework}{} \colorbox{mypink!55!white}{Bottom three scoring framework}}
\end{tabular}

\end{table}

\section{\rqref{rq:safeguards} \Revision{Mitigation Tasks}}\label{SEC:RQ2}

\textbf{\A{rq:safeguards-text}}
\MandatoryRevision{We answer RQ2 by analyzing mitigation tasks from Sec.~\ref{SEC:Metho-Val} and framework hit-count and coverage from Sec.~\ref{SEC:Metho-Measure}.
We detail mitigation tasks, starter kit tasks, and framework hit-count and coverage.}

\Revision{\textbf{Mitigation tasks.} We found 37 tasks that mitigate attack techniques of the \gls{aus}.
Out of the 37 mitigating tasks, 34 are mapped to P-SSCRM. 
We detail the missing mitigating tasks in Sec.~\ref{SEC:RQ3}.
Thus, only 34 out of the 73 P-SSCRM tasks mitigate attack techniques of the \gls{aus}. 
Out of the 37 mitigation tasks, 37 tasks mitigate attack techniques in \gls{solarwinds}, 34 mitigate \gls{log4j}, and 29 mitigate \gls{xz}.
Fig.~\ref{fig:task-overlap} shows the overlap of the mitigation tasks across \gls{aus}.}
\Revision{Only 27 tasks mitigate attack techniques in all 3 \gls{aus}.} 
Also, the \gls{solarwinds} tasks are a superset of the \gls{log4j} and \gls{xz} tasks.
\Revision{Fig.~\ref{fig:relationship} shows the relationship between \gls{psscrm} groups and practices for the attack techniques of each \gls{aus}.}
\Revision{Tasks of all four \gls{psscrm} groups mitigate attack techniques of the \gls{aus}}.
The environment (E) is the group with the highest number of attack techniques mitigated by tasks due to the secure software development environment (E.3) practice.
The second highest group is the deployment (D), mainly due to the monitoring intrusions or violations (D.2) practice.
We found mitigation tasks in 14 out of the 15 practices, excluding the developing security requirements (P.1) practice.

\Revision{\textbf{Starter kit tasks.}} The mitigating tasks with the top ten highest scores for the \gls{starter} are shown in Table~\ref{tab:priority}.
The top five \gls{starter} tasks are:
role-based access control (E.3.3)
system monitoring (D.2.1),
boundary protection (E.3.7),
monitor changes to configuration settings (E.3.6), and
environmental scanning tools (E.3.11).
Four of the top five \gls{starter} tasks are within the software development environment (E.3) practice.
The bottom five \gls{starter} tasks are: 
security design review (P.2.1),
dependency update (P.5.2),  
information flow enforcement (E.3.4),
protect information at rest (G.2.6), and
risk-based vulnerability remediation (D.1.2).
For each task, we use the attack techniques associated adversarial event compromised component to determine whether the mitigation applied to the dependent, dependency, or both.
All ten \gls{starter} tasks mitigate attack techniques at both the dependency and dependent levels.

\MandatoryRevision{\textbf{Framework hit-count and coverage.}} 
\MandatoryRevision{The framework hit-count and coverage is shown in Table~\ref{tab:Frameworks}.
We ordered the frameworks based on the number of times the framework is in the top-three minus the bottom-three for coverage and completeness.
For example, \acrshort{scorecard} is in two top-three scoring frameworks and three bottom-three, hence the difference is $2 - 3 =-1$.
In case of a tie, the average of the framework measures is used.}
\MandatoryRevision{
A higher hit count indicates that more attack techniques are covered by the tasks of the framework.
The referenced framework with the highest hit count was \gls{nist} 800-161, while the lowest was \gls{slsa}.
On average, frameworks mitigate 25.1 attack techniques out of the 203 in \gls{attack}.}

\Revision{Meanwhile,} higher coverage indicates that when adopting all tasks in a framework, more mitigation tasks are adopted.
Thus, based upon our results, software organizations are adopting fewer unused tasks.
\Gls{nist} 800-161 and \acrshort{attestation} for all three \gls{aus} have one of the three highest coverages.
\Revision{Meanwhile, \gls{cncf} SSC and \gls{s2c2f} have the lowest coverage overall for the \gls{aus}, ranking for two \gls{aus} in the bottom-three.}
No framework has coverage that exceeds 67\% for any \gls{aus}.
\Revision{Hence, based on our results, at least a third of the tasks are unused for the \gls{aus}.}
The \gls{aus} with the highest framework coverage is \gls{solarwinds}, followed by \gls{log4j}, and finally \gls{xz}.
\MandatoryRevision{Additionally, hit-count and framework coverage provided different rankings of the frameworks.
For example, \gls{cncf} SSC has the second highest hit-count of the frameworks, yet is of the lowest ranking in coverage.
Hence, frameworks that mitigate more attack techniques do not necessarily mitigate more attack techniques used in the \gls{aus}.}

\begin{tcolorbox}[myhighlightbox]
    \Revision{
    \textbf{Key Insights:} Less than half of the referenced frameworks' tasks mitigate AUS attack techniques (34 out of 73 tasks).
    The top three mitigation tasks in the starter kit are role-based access control, system monitoring, and boundary protection.
    }
\end{tcolorbox}

\section{\rqref{rq:frameworks} \Revision{Missing Tasks}}\label{SEC:RQ3}

\textbf{\A{rq:frameworks-text}}
\Revision{We answer RQ3 by analyzing the task gaps from Sec.~\ref{SEC:Metho-Gap} and framework completeness from Sec.~\ref{SEC:Metho-Measure}.
We describe framework completeness, \gls{psscrm} gaps, and industry-wide tasks.}

\Revision{\textbf{Framework completeness.}}
The referenced frameworks' completeness for each \gls{aus} is shown in Table~\ref{tab:Frameworks}.
Higher completeness indicates that when adopting all tasks in a framework, there are fewer missing mitigation tasks.
Hence, organizations would be less vulnerable to software supply chain attacks when implementing all the tasks in the framework.
\Revision{\Gls{nist} 800-161 and \gls{eo}/\gls{ssdf} have one of the highest completeness for the three \gls{aus}.}
\Revision{\acrshort{owaspSCVS}, \acrshort{scorecard}, and \gls{slsa} have one of the lowest completeness scores across all three \gls{aus}.}
Out of the referenced frameworks, only \Gls{nist} 800-161 has scored above or equal to 68\% for the three \gls{aus}.
Five of the ten referenced frameworks have an average completeness score below 40\%. 
\Revision{Hence, half referenced frameworks are missing more than 60\% of mitigating tasks on average.}
The average completeness differs for less than 2\% between the \gls{aus}.

\textbf{Gaps in \gls{psscrm}.} 
As \gls{psscrm} is the union of the ten referenced frameworks, any missing task is a gap across all the referenced frameworks.
Thus, organizations would still be vulnerable to software supply chain attacks even after adopting all tasks enumerated in \gls{psscrm}. 
We classified each gap we found into a full, major, or minor gap with the referenced \gls{cti} reports available and description in the supplemental material~\cite{supplementalLink}.

We found three \textit{full gaps}, new tasks, in \gls{psscrm} and we created new identifiers for each. 
The completeness of \gls{psscrm} ranges from 90\% to 92\%, as shown in Table~\ref{tab:Frameworks}.
The full gaps are as follows: 
\begin{itemize}[leftmargin=*]
    \item \textit{Sustainable open-source software} (G.5.5). Open-source is not a supplier, as no contractual relationship exists between dependencies and dependents. 
    Thus, security cannot be shifted as the responsibility of open-source, and instead, organizations should be responsible consumers and sustainable contributors.
    Mechanisms mentioned included donating, paying open-source developers more, channeling funds to projects, fundraising for regular audits of critical projects, and having paid staff.
    \item \textit{Environmental scanning tools} (E.3.11). Automated scanning tools can detect anomalies and undesired behaviors in each environment, including development, build, testing, and deployment. For example, tools to remove viruses, scan files downloaded from the internet before execution, detect signatures of anomalous payloads, and identify changes in configuration settings. The full gap is within the \gls{starter} tasks.
    \item \textit{Response partnerships} (D.1.7). Partners can help during vulnerability response by providing information about compromises, enabling improved response efforts, and collaborating in dealing with attacks. Adversaries are only empowered when no collaboration exists between software organizations.
\end{itemize}

We only found one \textit{major gap}, expanding the scope of an existing task, in \gls{psscrm} for dependency updates (P.5.2).
The task did not include staging or flighting releases when adopting component changes. 
The task was notably used in \gls{xz} in some distributions, such as Debian, reducing the number of affected dependent systems and is in the \gls{starter}.
Finally, we also updated five tasks from the \textit{minor gaps}, improving the tasks descriptions.
For example, specifying firewalls in boundary protection (E.3.7).
Three of the five tasks with minor gaps are in the \gls{starter}.

\textbf{Industry-wide tasks.} 
We found seven industry-wide efforts to help the software supply chain community mentioned in the \gls{cti} reports we analyzed (Sec.~\ref{SEC:Metho-Map}) that fall outside the scope of \gls{psscrm} to provide organizations with proactive security risk reduction tasks.
The task are identified in the form I.X. 
The \gls{cti} report references available in the supplemental material~\cite{supplementalLink} and the industry-wide tasks are explained below:

\begin{itemize}[leftmargin=*]
    \item \textit{Create and join task forces} (I.1). When high-profile incidents are discovered, the effort, resources, policies, collaboration, and culture required to eradicate the vulnerability transcend an organization. Unified task forces, such as the US Cyber Safety Review Board and OpenSSF, that have been created can aid.
    \item \textit{Create and evolve frameworks} (I.2). Frameworks for software supply chain security, such as the ten referenced frameworks in \gls{psscrm}, should be created to aid organizations. At the same time, current frameworks should continue to evolve, be extended, and improved in new versions.
    Framework adoption should be practical and encouraging.
    \item \textit{Create and improve tools} (I.3). The industry should create and improve tools to reduce the manual effort required in detecting vulnerabilities, changing dependencies, scanning weaknesses, increasing security guarantees provided, and measuring tasks. \Revision{The tools should also be usable and acceptable for adopters.}
    \item \textit{Create and improve regulations} (I.4). \Revision{Regulations should be created and extended to cover software supply chain security that incentivizes good actions.} For example, prompt disclosure should be incentivized rather than sanctioning companies disclosing vulnerabilities such as in \gls{log4j}.
    \item \textit{Expand investments} (I.5). Education, government, and research investments should be expanded. The investments include improving curricula, integrating training during degrees, providing grants, funding projects, and researching novel approaches.
    \item \textit{Analyze incidents} (I.6). Without learning from the attacks, past mistakes leading to the software supply chain incidents are bound to be repeated. Therefore, the industry can reflect on opportunities to improve the security posture.
    \item \textit{Foster a supportive community} (I.7). Adversaries prey upon developers through social engineering attacks, while victims targeted are subjected to scrutiny, speculation, increased demands, and professional burnout.
Software supply chain security cannot sustain a model without fostering a community prioritizing mental health, empathy, and members' long-term care. 
\end{itemize}

\begin{tcolorbox}[myhighlightbox]
    \Revision{
    \textbf{Key Insights:} We found three missing tasks in P-SSCRM and, therefore, in all ten contributing frameworks. 
    Hence, products of software organizations would still be vulnerable to attacks even if adopting all tasks in the ten referenced frameworks.
    }
\end{tcolorbox}

\section{Discussion}\label{sec:discussion}

Based on our findings, we provide the following recommendations for software organizations, framework authors, and researchers.

\textbf{Software organizations should prioritize adopting \gls{starter} tasks.} 
\Revision{In Sec.~\ref{SEC:RQ2}, we found the top ten mitigation tasks we gathered in a \gls{starter}. We recommend that software organizations prioritize adopting the tasks based on the ranking.}
\Revision{Notably, open source software projects that lack time and resources.}

Interestingly, most of the \gls{starter} tasks apply to broader software security rather than being specific to the software supply chain security and we hypothesize two reasons.
First, software security and software supply chain security risks are similar.
For the same reason, \gls{starter} tasks are related to the OWASP Top 2021 categories~\cite{owasptop}. 
For example, role-based access control (E.3.3) with broken access control (A01:2021), system monitoring (D.2.1) with security logging and monitoring failures (A09:2021), and dependency update (P.5.2) with vulnerable and outdated components (A06:2021).
Second, before mitigating software supply chain attacks, common software security tasks should be addressed.
Therefore, software supply chain specific tasks are aspirational and forward-looking.
Additionally, as we select tasks that mitigate attack techniques, the tasks we find are also technical.
Hence, tasks that support the entire software supply chain effort, likely in the governance (G) \gls{psscrm} group, are not found in our approach.

Another intriguing observation is that we found tasks from all four \gls{psscrm} groups in the \gls{starter}.
Thus, software supply chain security also requires defense-in-depth.
\Revision{For the same reason, referenced frameworks with higher coverage and completeness did not have a ``weak link group'' with few tasks.}
For example, each \gls{psscrm} group in  \acrshort{attestation}, one of the three frameworks with no bottom-three score, has at least four tasks (Table~\ref{TAB:PSSCRM}).
Software organizations can consider the coverage and completeness measures when selecting a framework.
Organizations starting to adopt referenced frameworks should focus on coverage to use resources more effectively.
Meanwhile, more mature organizations should focus on completeness to reduce gaps within the frameworks and use a union of frameworks to minimize each gap.
\Revision{Finally, platforms and ecosystems should support and enable prioritized tasks, similar to how GitHub enforces Multi-Factor Authentication. }

\textbf{Framework authors should add appropriate mitigation tasks to improve the frameworks.}
\Revision{In Sec.~\ref{SEC:RQ2} and~\ref{SEC:RQ3}, we found mitigation and gaps tasks for the \gls{aus}. 
We recommend that each framework add appropriate mitigation tasks, based on the framework's focus shown in Table~\ref{TAB:PSSCRM}.
We take into consideration such differences in our following recommendations.}

For \gls{slsa}, the low completeness is due to \gls{slsa} v1.0 removing tasks~\cite{slsaUpdate}.
Additionally, as the authors plan to align and combine \gls{slsa} and \gls{s2c2f}~\cite{slsaNextVersion}, the coverage and completeness of the updated \gls{slsa} should improve.
We additionally recommend, in line with \gls{slsa} focus of securing the build integrity, to also include the secured orchestration platform (E.2.5) task.
Due to the low completeness, we recommend including and researching more automated measures in \acrshort{scorecard}, in line with the referenced framework use case.
Still, as the framework also achieves higher coverage, at least 44\% of tasks mitigate attack techniques in the \gls{aus}.
Meanwhile, \gls{nist} 800-161 achieved the highest coverage due to the framework scope, including manufacturing and hardware.
\Revision{Lastly, as US NIST \gls{ssdf} is a set of fundamental software security tasks~\cite{ssdf}, we recommend US NIST \gls{ssdf} to incorporate the top five \gls{starter} tasks, as none are included in the framework.}

\Revision{We found three full gaps in \gls{psscrm}, which are being integrated by the authors in an update.}
\MandatoryRevision{Interestingly, the gaps we found have been investigated outside of the analyzed reports from which we collected the gaps.
For example, environmental scanning tools have been studied~\cite {lin2024untrustide}.}
\MandatoryRevision{Hence, task gaps are not software security knowledge gaps, but rather gaps in the referenced frameworks.}
Additionally, for the industry-wide tasks out of scope for \gls{psscrm}, an I-SSCRM (industry-wide) framework should be created to centralize efforts.
Future research should study and expand tasks in I-SSCRM.
The community should also be involved with I-SSCRM tasks that are transversal and support \gls{psscrm} tasks.
Still, prior work has found that getting industry-wide participation is one of the top five challenges for software supply chain security~\cite{enck2022top}, hence such effort is not trivial and will require time.

\textbf{Researchers should continue gathering threat intelligence trends for software supply chain attacks.}
In Sec.~\ref{SEC:RQ1}, we found the attack techniques used in each \gls{aus}.
\Revision{We recommend that researchers focus on mitigating the 12 attack techniques that overlap between the AUS and leveraging our insights about the attacks.}

\Revision{Yet, more research is required.}
Attacks should continue to be analyzed, particularly high-profile incidents.
\Revision{As frameworks tasks will evolve and new notable attacks emerge, we plan to annually replicate our methodology to update the \gls{starter} tasks. 
We will also include in a future iteration more measurement approaches, such as MITRE's Top ATT\&CK Techniques~\cite{mitreATTACKTop}.
We have also shared in supplemental material our analysis scripts to enable others to replicate and extend our research~\cite{supplementalLink}.}
Second, research should automate software supply chain threat intelligence insights.
We decided for our research design to manually analyze the events in the \gls{aus} to have granular and high-quality data. 
Still, future work can use existing attack technique-level automated techniques~\cite{rahman2023attackers}, automate event-level analysis, or explore other approaches.
Additionally, work can leverage the attack technique to task mappings we constructed~\cite{supplementalLink}.
\Revision{\gls{cti} reports should also provide explicit attack techniques mappings and frameworks attack technique to task mappings to ease analysis.}
\Revision{Finally, our} work provides a foundation for future work to investigate further the attacks, including researching why the events happened, what alternative attack techniques adversaries could have employed, or how effective the tasks are.

\section{Limitations}
\label{sec:threats}

The following are the limitations of our study.

\textbf{Generalizability.} 
The findings of the software supply chain attacks are limited to the \gls{aus}.
However, as we deliberately choose the \gls{aus}, we account for the limitation by selecting representative high-profile examples for each main software supply chain attack vector~\cite{williams2024research}. 
Another limitation is the generalizability outside of the \gls{cti} reports gathered.
\MandatoryRevision{We mitigate the threat by using three independent sampling strategies to find reports (Sec.~\ref{SEC:Metho-Col}) and achieve saturation after analyzing the G3 reports (Sec.~\ref{SEC:Metho-Qual}).}

\textbf{Interpretation.} 
\Revision{As we qualitatively analyzed the CTI reports, we are subject to and acknowledge biases in interpretation.}
At least two researchers were involved in the qualitative analysis to reduce biases until agreement was achieved.
\MandatoryRevision{We state the authors' expertise to increase transparency (Sec.~\ref{sec:metho}).}
\Revision{Finally, we share our data and scripts to improve transparency and adoption~\cite{supplementalLink}.}

\textbf{Credibility.}
\Revision{Consistent with prior work~\cite{chen2024contents}, we rely on the quality and credibility of the CTI reports analyzed. We mitigate the threat by sampling reports from expert sources (Sec.~\ref{SEC:Metho-Col}).}
Creating a credible attack technique to task mapping at scale is subject to biases. 
\Revision{We mitigate biases by using four independent strategies to gather candidate mappings to provide diverse opinions and find convergence (Sec.~\ref{SEC:Metho-Map}).}
Other candidate mapping strategies approaches may be integrated to extend our work.

\section{Related Work}
\label{sec:relwork}

\textbf{Software supply chain tasks and frameworks.}
\Revision{Software supply chain research has studied the practitioner task adoption challenges~\cite{fourne2023s, stalnaker2024boms, kalu2024industry}, tasks adoption~\cite{benedettiempirical}, tooling deficiencies~\cite{yu2024correctness}, and tasks effectiveness~\cite{he2025pinning}.}
Tools to implement tasks, such as \textit{sigstore}~\cite{newman2022sigstore} and \textit{in-toto}~\cite{torres2019toto}, have also been proposed.
The most similar work to ours is by Zahan et al.~\cite{zahan2023practices} that prioritized \acrshort{scorecard} tasks based on security outcomes.
\Revision{Instead, we create a \gls{starter} that ranks the \gls{psscrm} tasks, \Revision{which} include \acrshort{scorecard}, based on the mitigated attack techniques from three \gls{aus}\Revision{.}}
Furthermore, research has also systematized and categorized the software supply chain in mappings between: attack vectors to tasks~\cite{ladisa2023sok}, security properties to frameworks~\cite{okafor2022sok}, and core software supply chain elements to tasks~\cite{ishgair2024sok}.
Additionally, bi-directionally equivalent tasks of frameworks have been mapped~\cite{williams2024proactive}.
Our work leverages and builds upon prior work.
Yet, we differ as we map attacks to attack techniques to provide more granular detail of the attacks, we then systematically map to mitigation framework tasks.

\noindent \textbf{Software security incident analysis.}
Software security research has investigated attacks.
\MandatoryRevision{Particularly, the three \gls{aus} have been priorly investigated~\cite{chowdhury2022better,peisert2021perspectives,lins2024critical,sarang2024plotting,hiesgen2024log4j, przymus2025wolves}.}
\MandatoryRevision{
Yet, few studies have focused on collecting the attack techniques leveraged.
We found only 6 out of 106 \gls{cti} reports explicitly indicating attack techniques for the \gls{solarwinds} and \gls{log4j} incidents.
Notably, MITRE~\cite{mitreSWCampaign} and CISA~\cite{cisa2021SolarWinds,cisa2021APT} synthesized reports of the \gls{solarwinds} incident.
Our work builds upon such work, combining all 106 \gls{cti} reports to strive for theoretical saturation of all the \gls{aus}.
Furthermore, we enhance the synthesis by mapping the \gls{aus} to framework tasks with an evidence-based \gls{starter} and evaluating the frameworks.
}

\noindent \Revision{\textbf{Software security framework evaluation.}}
\Revision{
Prior work has also evaluated the content of security frameworks~\cite{stevens2020compliance, stevens2022ready, stevens2020lurks}.
The most similar work is by Chen et al.~\cite{chen2024contents} investigating Internet of Things frameworks, including a retrospective evaluation of vulnerability prevention.
In line with prior work, we also evaluate software security frameworks, specifically software supply chain frameworks.
We further enhance prior work by providing a novel, granular evaluation procedure for software security frameworks through the mappings between attack techniques and tasks.
}

\section{Conclusion}
\label{sec:conc}

\Revision{With the rise of software supply chain attacks, frameworks have been developed to provide \textit{what tasks software organizations should or must adopt} to reduce software security risk.}
\Revision{Still, software organizations would also benefit from knowing \textit{what tasks mitigate attack techniques} currently used by attackers.}
\MandatoryRevision{In this empirical study, we systematically synthesize how framework tasks mitigate the attack techniques in the \gls{solarwinds}, \gls{log4j}, and \gls{xz} attacks. 
We analyzed 106 \gls{cti} reports, mapped reports to 203 attack techniques, mapped attack techniques to 73 tasks, identified framework gaps from reports, and established a ranked \gls{starter} for organizations.}
\Revision{We found that across all \Revision{attacks}, there is a common set of 12 attack techniques.
\Revision{The top ten mitigating tasks in the \gls{starter} included} role-based access control (E.3.3), system monitoring (D.2.1), and boundary protection (E.3.7).
We also found \Revision{three} missing tasks in \gls{psscrm} and thus in the \Revision{ten} referenced frameworks.}
Thus, software products would still be vulnerable to software supply chain attacks even if organizations adopted all recommended tasks.

\begin{acks}
The work was supported and funded by the National Science Foundation Grant No. 2207008 and the North Carolina State University Goodnight Doctoral Fellowship. 
Any opinions expressed in this material are those of the authors and do not necessarily reflect the views of any of the funding organizations. 
We profoundly thank the Realsearch, WSPR, and S3C2 research group members, visitors, and collaborators for their support and feedback.
\end{acks}

\bibliographystyle{ACM-Reference-Format}
\bibliography{references}

\clearpage
\onecolumn
\appendix
\section{MITRE ATT\&CK attack techniques to P-SSCRM tasks mappings}
We \textbf{bold} the rows for the \glspl{attack} to \glspl{safeguard} mappings found by all four strategies. We \textit{italicize} rows related to the full gaps.
\begin{longtable}{llllll}
\toprule
MITRE ATT\&CK Technique & P-SSCRM Task & Transitive (M1) & LLM (M2) & Framework (M3) & Report (M4) \\
\midrule
\endfirsthead
\toprule
MITRE ATT\&CK Technique & P-SSCRM Task & Transitive (M1) & LLM (M2) & Framework (M3) & Report (M4) \\
\midrule
\endhead
\midrule
\multicolumn{6}{r}{Continued on next page} \\
\midrule
\endfoot
\bottomrule
\endlastfoot
T1190 & D.1.2 & $\checkmark$ & $\checkmark$ &  & $\checkmark$ \\
T1195 & D.1.2 & $\checkmark$ & $\checkmark$ &  & $\checkmark$ \\
T1553 & D.1.2 & $\checkmark$ & $\checkmark$ &  & $\checkmark$ \\
T1562 & D.1.2 & $\checkmark$ & $\checkmark$ &  & $\checkmark$ \\
T1190 & D.1.3 & $\checkmark$ & $\checkmark$ &  & $\checkmark$ \\
T1195 & D.1.3 & $\checkmark$ & $\checkmark$ &  & $\checkmark$ \\
T1195 & D.1.4 & $\checkmark$ & $\checkmark$ &  & $\checkmark$ \\
T1553 & D.1.4 & $\checkmark$ & $\checkmark$ &  & $\checkmark$ \\
T1195 & D.1.6 & $\checkmark$ & $\checkmark$ &  & $\checkmark$ \\
T1003 & D.2.1 & $\checkmark$ & $\checkmark$ &  & $\checkmark$ \\
T1021 & D.2.1 &  & $\checkmark$ & $\checkmark$ & $\checkmark$ \\
T1027 & D.2.1 &  & $\checkmark$ & $\checkmark$ & $\checkmark$ \\
T1036 & D.2.1 & $\checkmark$ &  & $\checkmark$ & $\checkmark$ \\
T1041 & D.2.1 & $\checkmark$ & $\checkmark$ &  & $\checkmark$ \\
T1068 & D.2.1 & $\checkmark$ & $\checkmark$ & $\checkmark$ &  \\
T1070 & D.2.1 & $\checkmark$ & $\checkmark$ &  & $\checkmark$ \\
T1071 & D.2.1 & $\checkmark$ & $\checkmark$ &  & $\checkmark$ \\
T1078 & D.2.1 & $\checkmark$ & $\checkmark$ &  & $\checkmark$ \\
T1080 & D.2.1 & $\checkmark$ & $\checkmark$ & $\checkmark$ &  \\
T1105 & D.2.1 & $\checkmark$ & $\checkmark$ &  & $\checkmark$ \\
T1176 & D.2.1 & $\checkmark$ & $\checkmark$ & $\checkmark$ &  \\
T1189 & D.2.1 & $\checkmark$ & $\checkmark$ & $\checkmark$ &  \\
T1195 & D.2.1 & $\checkmark$ & $\checkmark$ &  & $\checkmark$ \\
T1203 & D.2.1 & $\checkmark$ & $\checkmark$ & $\checkmark$ &  \\
T1210 & D.2.1 & $\checkmark$ & $\checkmark$ & $\checkmark$ &  \\
T1211 & D.2.1 & $\checkmark$ & $\checkmark$ & $\checkmark$ &  \\
T1212 & D.2.1 & $\checkmark$ & $\checkmark$ & $\checkmark$ &  \\
T1213 & D.2.1 & $\checkmark$ & $\checkmark$ & $\checkmark$ &  \\
T1484 & D.2.1 &  & $\checkmark$ & $\checkmark$ & $\checkmark$ \\
T1528 & D.2.1 & $\checkmark$ & $\checkmark$ & $\checkmark$ &  \\
T1543 & D.2.1 & $\checkmark$ & $\checkmark$ & $\checkmark$ &  \\
T1550 & D.2.1 &  & $\checkmark$ & $\checkmark$ & $\checkmark$ \\
T1555 & D.2.1 & $\checkmark$ & $\checkmark$ &  & $\checkmark$ \\
T1556 & D.2.1 & $\checkmark$ & $\checkmark$ & $\checkmark$ &  \\
T1568 & D.2.1 & $\checkmark$ & $\checkmark$ &  & $\checkmark$ \\
T1573 & D.2.1 & $\checkmark$ &  & $\checkmark$ & $\checkmark$ \\
T1574 & D.2.1 & $\checkmark$ & $\checkmark$ & $\checkmark$ &  \\
T1606 & D.2.1 &  & $\checkmark$ & $\checkmark$ & $\checkmark$ \\
T1078 & E.1.3 &  & $\checkmark$ & $\checkmark$ & $\checkmark$ \\
T1098 & E.1.3 &  & $\checkmark$ & $\checkmark$ & $\checkmark$ \\
T1072 & E.1.5 & $\checkmark$ & $\checkmark$ &  & $\checkmark$ \\
T1190 & E.2.4 & $\checkmark$ &  & $\checkmark$ & $\checkmark$ \\
T1068 & E.2.5 & $\checkmark$ & $\checkmark$ & $\checkmark$ &  \\
T1072 & E.2.5 & $\checkmark$ & $\checkmark$ &  & $\checkmark$ \\
T1190 & E.2.5 & $\checkmark$ & $\checkmark$ & $\checkmark$ &  \\
T1211 & E.2.5 & $\checkmark$ & $\checkmark$ & $\checkmark$ &  \\
T1212 & E.2.5 & $\checkmark$ & $\checkmark$ & $\checkmark$ &  \\
T1040 & E.3.2 & $\checkmark$ & $\checkmark$ & $\checkmark$ &  \\
T1072 & E.3.2 &  & $\checkmark$ & $\checkmark$ & $\checkmark$ \\
T1078 & E.3.2 & $\checkmark$ & $\checkmark$ &  & $\checkmark$ \\
T1557 & E.3.2 & $\checkmark$ & $\checkmark$ & $\checkmark$ &  \\
T1563 & E.3.2 & $\checkmark$ & $\checkmark$ & $\checkmark$ &  \\
T1003 & E.3.3 & $\checkmark$ & $\checkmark$ & $\checkmark$ &  \\
T1021 & E.3.3 & $\checkmark$ & $\checkmark$ & $\checkmark$ &  \\
T1036 & E.3.3 & $\checkmark$ &  & $\checkmark$ & $\checkmark$ \\
T1047 & E.3.3 & $\checkmark$ & $\checkmark$ & $\checkmark$ &  \\
T1059 & E.3.3 & $\checkmark$ &  & $\checkmark$ & $\checkmark$ \\
T1068 & E.3.3 & $\checkmark$ & $\checkmark$ &  & $\checkmark$ \\
T1080 & E.3.3 & $\checkmark$ & $\checkmark$ & $\checkmark$ &  \\
T1110 & E.3.3 & $\checkmark$ & $\checkmark$ & $\checkmark$ &  \\
T1134 & E.3.3 & $\checkmark$ & $\checkmark$ & $\checkmark$ &  \\
T1136 & E.3.3 & $\checkmark$ & $\checkmark$ & $\checkmark$ &  \\
T1185 & E.3.3 & $\checkmark$ & $\checkmark$ & $\checkmark$ &  \\
T1210 & E.3.3 & $\checkmark$ & $\checkmark$ & $\checkmark$ &  \\
T1213 & E.3.3 & $\checkmark$ & $\checkmark$ & $\checkmark$ &  \\
T1218 & E.3.3 & $\checkmark$ & $\checkmark$ & $\checkmark$ &  \\
T1222 & E.3.3 & $\checkmark$ & $\checkmark$ & $\checkmark$ &  \\
T1484 & E.3.3 & $\checkmark$ & $\checkmark$ & $\checkmark$ &  \\
T1489 & E.3.3 & $\checkmark$ & $\checkmark$ & $\checkmark$ &  \\
T1505 & E.3.3 & $\checkmark$ & $\checkmark$ & $\checkmark$ &  \\
T1525 & E.3.3 & $\checkmark$ & $\checkmark$ & $\checkmark$ &  \\
T1528 & E.3.3 & $\checkmark$ & $\checkmark$ & $\checkmark$ &  \\
T1530 & E.3.3 & $\checkmark$ & $\checkmark$ & $\checkmark$ &  \\
T1537 & E.3.3 & $\checkmark$ & $\checkmark$ & $\checkmark$ &  \\
T1538 & E.3.3 & $\checkmark$ & $\checkmark$ & $\checkmark$ &  \\
T1550 & E.3.3 & $\checkmark$ & $\checkmark$ & $\checkmark$ &  \\
T1553 & E.3.3 & $\checkmark$ & $\checkmark$ & $\checkmark$ &  \\
T1556 & E.3.3 & $\checkmark$ & $\checkmark$ & $\checkmark$ &  \\
T1558 & E.3.3 & $\checkmark$ & $\checkmark$ & $\checkmark$ &  \\
T1563 & E.3.3 & $\checkmark$ & $\checkmark$ & $\checkmark$ &  \\
T1565 & E.3.3 & $\checkmark$ & $\checkmark$ & $\checkmark$ &  \\
T1569 & E.3.3 & $\checkmark$ & $\checkmark$ & $\checkmark$ &  \\
T1578 & E.3.3 & $\checkmark$ & $\checkmark$ & $\checkmark$ &  \\
T1609 & E.3.3 & $\checkmark$ & $\checkmark$ & $\checkmark$ &  \\
T1610 & E.3.3 & $\checkmark$ & $\checkmark$ & $\checkmark$ &  \\
T1621 & E.3.3 & $\checkmark$ & $\checkmark$ & $\checkmark$ &  \\
T1649 & E.3.3 & $\checkmark$ & $\checkmark$ & $\checkmark$ &  \\
T1651 & E.3.3 & $\checkmark$ & $\checkmark$ & $\checkmark$ &  \\
T1040 & E.3.4 & $\checkmark$ & $\checkmark$ & $\checkmark$ &  \\
T1114 & E.3.4 & $\checkmark$ & $\checkmark$ & $\checkmark$ &  \\
T1213 & E.3.4 & $\checkmark$ & $\checkmark$ & $\checkmark$ &  \\
T1530 & E.3.4 & $\checkmark$ & $\checkmark$ & $\checkmark$ &  \\
T1552 & E.3.4 & $\checkmark$ & $\checkmark$ & $\checkmark$ &  \\
T1557 & E.3.4 & $\checkmark$ & $\checkmark$ & $\checkmark$ &  \\
T1565 & E.3.4 & $\checkmark$ & $\checkmark$ & $\checkmark$ &  \\
T1602 & E.3.4 & $\checkmark$ & $\checkmark$ & $\checkmark$ &  \\
T1659 & E.3.4 & $\checkmark$ & $\checkmark$ & $\checkmark$ &  \\
T1059 & E.3.6 & $\checkmark$ &  & $\checkmark$ & $\checkmark$ \\
T1098 & E.3.6 & $\checkmark$ & $\checkmark$ &  & $\checkmark$ \\
T1114 & E.3.6 & $\checkmark$ &  & $\checkmark$ & $\checkmark$ \\
T1176 & E.3.6 & $\checkmark$ & $\checkmark$ & $\checkmark$ &  \\
T1505 & E.3.6 & $\checkmark$ & $\checkmark$ & $\checkmark$ &  \\
T1525 & E.3.6 & $\checkmark$ & $\checkmark$ & $\checkmark$ &  \\
T1543 & E.3.6 & $\checkmark$ & $\checkmark$ & $\checkmark$ &  \\
T1548 & E.3.6 & $\checkmark$ & $\checkmark$ & $\checkmark$ &  \\
T1550 & E.3.6 & $\checkmark$ &  & $\checkmark$ & $\checkmark$ \\
T1552 & E.3.6 & $\checkmark$ &  & $\checkmark$ & $\checkmark$ \\
T1556 & E.3.6 & $\checkmark$ & $\checkmark$ & $\checkmark$ &  \\
T1562 & E.3.6 & $\checkmark$ & $\checkmark$ & $\checkmark$ &  \\
T1574 & E.3.6 & $\checkmark$ & $\checkmark$ & $\checkmark$ &  \\
T1610 & E.3.6 & $\checkmark$ & $\checkmark$ & $\checkmark$ &  \\
T1612 & E.3.6 & $\checkmark$ & $\checkmark$ & $\checkmark$ &  \\
T1001 & E.3.7 & $\checkmark$ &  & $\checkmark$ & $\checkmark$ \\
T1008 & E.3.7 & $\checkmark$ & $\checkmark$ & $\checkmark$ &  \\
T1021 & E.3.7 &  & $\checkmark$ & $\checkmark$ & $\checkmark$ \\
T1047 & E.3.7 &  & $\checkmark$ & $\checkmark$ & $\checkmark$ \\
T1048 & E.3.7 & $\checkmark$ & $\checkmark$ & $\checkmark$ &  \\
T1072 & E.3.7 & $\checkmark$ & $\checkmark$ & $\checkmark$ &  \\
T1080 & E.3.7 & $\checkmark$ & $\checkmark$ & $\checkmark$ &  \\
T1090 & E.3.7 & $\checkmark$ & $\checkmark$ & $\checkmark$ &  \\
T1095 & E.3.7 & $\checkmark$ & $\checkmark$ & $\checkmark$ &  \\
T1102 & E.3.7 & $\checkmark$ &  & $\checkmark$ & $\checkmark$ \\
T1104 & E.3.7 & $\checkmark$ & $\checkmark$ & $\checkmark$ &  \\
T1105 & E.3.7 & $\checkmark$ & $\checkmark$ & $\checkmark$ &  \\
T1133 & E.3.7 & $\checkmark$ & $\checkmark$ & $\checkmark$ &  \\
T1176 & E.3.7 & $\checkmark$ & $\checkmark$ & $\checkmark$ &  \\
T1187 & E.3.7 & $\checkmark$ & $\checkmark$ & $\checkmark$ &  \\
T1189 & E.3.7 & $\checkmark$ & $\checkmark$ & $\checkmark$ &  \\
T1199 & E.3.7 & $\checkmark$ & $\checkmark$ &  & $\checkmark$ \\
T1203 & E.3.7 & $\checkmark$ & $\checkmark$ & $\checkmark$ &  \\
T1204 & E.3.7 & $\checkmark$ & $\checkmark$ & $\checkmark$ &  \\
T1205 & E.3.7 & $\checkmark$ & $\checkmark$ & $\checkmark$ &  \\
T1210 & E.3.7 & $\checkmark$ & $\checkmark$ & $\checkmark$ &  \\
T1212 & E.3.7 & $\checkmark$ & $\checkmark$ & $\checkmark$ &  \\
T1219 & E.3.7 & $\checkmark$ & $\checkmark$ & $\checkmark$ &  \\
T1221 & E.3.7 & $\checkmark$ & $\checkmark$ & $\checkmark$ &  \\
T1537 & E.3.7 & $\checkmark$ & $\checkmark$ & $\checkmark$ &  \\
T1552 & E.3.7 & $\checkmark$ &  & $\checkmark$ & $\checkmark$ \\
T1557 & E.3.7 & $\checkmark$ & $\checkmark$ & $\checkmark$ &  \\
T1566 & E.3.7 & $\checkmark$ & $\checkmark$ & $\checkmark$ &  \\
T1567 & E.3.7 & $\checkmark$ & $\checkmark$ & $\checkmark$ &  \\
T1568 & E.3.7 & $\checkmark$ & $\checkmark$ & $\checkmark$ &  \\
T1570 & E.3.7 & $\checkmark$ & $\checkmark$ & $\checkmark$ &  \\
T1571 & E.3.7 & $\checkmark$ & $\checkmark$ & $\checkmark$ &  \\
T1572 & E.3.7 & $\checkmark$ & $\checkmark$ & $\checkmark$ &  \\
T1573 & E.3.7 & $\checkmark$ & $\checkmark$ & $\checkmark$ &  \\
T1599 & E.3.7 & $\checkmark$ & $\checkmark$ & $\checkmark$ &  \\
T1602 & E.3.7 & $\checkmark$ & $\checkmark$ & $\checkmark$ &  \\
T1610 & E.3.7 & $\checkmark$ & $\checkmark$ & $\checkmark$ &  \\
T1611 & E.3.7 & $\checkmark$ & $\checkmark$ & $\checkmark$ &  \\
T1659 & E.3.7 & $\checkmark$ & $\checkmark$ & $\checkmark$ &  \\
T1078 & E.3.8 &  & $\checkmark$ & $\checkmark$ & $\checkmark$ \\
T1552 & E.3.8 &  & $\checkmark$ & $\checkmark$ & $\checkmark$ \\
T1558 & E.3.8 &  & $\checkmark$ & $\checkmark$ & $\checkmark$ \\
T1078 & E.3.9 &  & $\checkmark$ & $\checkmark$ & $\checkmark$ \\
T1195 & G.1.4 & $\checkmark$ & $\checkmark$ &  & $\checkmark$ \\
T1195 & G.1.5 & $\checkmark$ & $\checkmark$ &  & $\checkmark$ \\
T1190 & G.2.5 & $\checkmark$ & $\checkmark$ &  & $\checkmark$ \\
T1005 & G.2.6 & $\checkmark$ & $\checkmark$ & $\checkmark$ &  \\
T1041 & G.2.6 & $\checkmark$ & $\checkmark$ & $\checkmark$ &  \\
T1048 & G.2.6 & $\checkmark$ & $\checkmark$ & $\checkmark$ &  \\
T1052 & G.2.6 & $\checkmark$ & $\checkmark$ & $\checkmark$ &  \\
T1530 & G.2.6 & $\checkmark$ & $\checkmark$ & $\checkmark$ &  \\
T1565 & G.2.6 & $\checkmark$ & $\checkmark$ & $\checkmark$ &  \\
T1195 & G.3.1 & $\checkmark$ & $\checkmark$ &  & $\checkmark$ \\
T1550 & G.3.4 &  & $\checkmark$ & $\checkmark$ & $\checkmark$ \\
T1606 & G.3.4 &  & $\checkmark$ & $\checkmark$ & $\checkmark$ \\
T1190 & G.4.1 & $\checkmark$ & $\checkmark$ &  & $\checkmark$ \\
T1195 & G.4.1 &  & $\checkmark$ & $\checkmark$ & $\checkmark$ \\
T1068 & G.4.3 & $\checkmark$ & $\checkmark$ & $\checkmark$ &  \\
T1195 & G.4.3 & $\checkmark$ & $\checkmark$ &  & $\checkmark$ \\
T1210 & G.4.3 & $\checkmark$ & $\checkmark$ & $\checkmark$ &  \\
T1211 & G.4.3 & $\checkmark$ & $\checkmark$ & $\checkmark$ &  \\
T1212 & G.4.3 & $\checkmark$ & $\checkmark$ & $\checkmark$ &  \\
T1190 & G.5.1 & $\checkmark$ & $\checkmark$ &  & $\checkmark$ \\
T1195 & G.5.1 & $\checkmark$ & $\checkmark$ &  & $\checkmark$ \\
T1199 & G.5.1 & $\checkmark$ & $\checkmark$ &  & $\checkmark$ \\
T1072 & G.5.2 & $\checkmark$ & $\checkmark$ &  & $\checkmark$ \\
T1046 & P.2.1 & $\checkmark$ &  & $\checkmark$ & $\checkmark$ \\
T1127 & P.2.1 & $\checkmark$ & $\checkmark$ & $\checkmark$ &  \\
T1195 & P.2.1 & $\checkmark$ & $\checkmark$ &  & $\checkmark$ \\
T1199 & P.2.1 & $\checkmark$ & $\checkmark$ &  & $\checkmark$ \\
T1210 & P.2.1 & $\checkmark$ & $\checkmark$ & $\checkmark$ &  \\
T1221 & P.2.1 & $\checkmark$ & $\checkmark$ & $\checkmark$ &  \\
T1505 & P.2.1 & $\checkmark$ & $\checkmark$ & $\checkmark$ &  \\
T1553 & P.2.1 & $\checkmark$ & $\checkmark$ &  & $\checkmark$ \\
T1559 & P.2.1 & $\checkmark$ & $\checkmark$ & $\checkmark$ &  \\
T1609 & P.2.1 & $\checkmark$ & $\checkmark$ & $\checkmark$ &  \\
T1611 & P.2.1 & $\checkmark$ & $\checkmark$ & $\checkmark$ &  \\
T1554 & P.3.3 &  & $\checkmark$ & $\checkmark$ & $\checkmark$ \\
T1195 & P.3.4 & $\checkmark$ & $\checkmark$ &  & $\checkmark$ \\
T1195 & P.4.1 & $\checkmark$ & $\checkmark$ &  & $\checkmark$ \\
T1190 & P.4.3 & $\checkmark$ & $\checkmark$ & $\checkmark$ &  \\
T1210 & P.4.3 & $\checkmark$ & $\checkmark$ & $\checkmark$ &  \\
T1190 & P.4.4 & $\checkmark$ & $\checkmark$ &  & $\checkmark$ \\
T1195 & P.4.5 & $\checkmark$ & $\checkmark$ &  & $\checkmark$ \\
T1505 & P.4.5 & $\checkmark$ & $\checkmark$ & $\checkmark$ &  \\
T1553 & P.4.5 & $\checkmark$ & $\checkmark$ &  & $\checkmark$ \\
T1612 & P.4.5 & $\checkmark$ & $\checkmark$ & $\checkmark$ &  \\
T1195 & P.5.1 & $\checkmark$ & $\checkmark$ &  & $\checkmark$ \\
T1068 & P.5.2 & $\checkmark$ & $\checkmark$ & $\checkmark$ &  \\
T1072 & P.5.2 & $\checkmark$ & $\checkmark$ & $\checkmark$ &  \\
T1210 & P.5.2 & $\checkmark$ & $\checkmark$ & $\checkmark$ &  \\
T1211 & P.5.2 & $\checkmark$ & $\checkmark$ & $\checkmark$ &  \\
T1212 & P.5.2 & $\checkmark$ & $\checkmark$ & $\checkmark$ &  \\
T1553 & P.5.2 & $\checkmark$ & $\checkmark$ &  & $\checkmark$ \\
\textbf{T1059} & \textbf{D.2.1} & $\checkmark$ & $\checkmark$ & $\checkmark$ & $\checkmark$ \\
\textbf{T1190} & \textbf{D.2.1} & $\checkmark$ & $\checkmark$ & $\checkmark$ & $\checkmark$ \\
\textbf{T1218} & \textbf{D.2.1} & $\checkmark$ & $\checkmark$ & $\checkmark$ & $\checkmark$ \\
\textbf{T1548} & \textbf{D.2.1} & $\checkmark$ & $\checkmark$ & $\checkmark$ & $\checkmark$ \\
\textbf{T1552} & \textbf{D.2.1} & $\checkmark$ & $\checkmark$ & $\checkmark$ & $\checkmark$ \\
\textbf{T1558} & \textbf{D.2.1} & $\checkmark$ & $\checkmark$ & $\checkmark$ & $\checkmark$ \\
\textbf{T1562} & \textbf{D.2.1} & $\checkmark$ & $\checkmark$ & $\checkmark$ & $\checkmark$ \\
\textbf{T1190} & \textbf{E.3.2} & $\checkmark$ & $\checkmark$ & $\checkmark$ & $\checkmark$ \\
\textbf{T1072} & \textbf{E.3.3} & $\checkmark$ & $\checkmark$ & $\checkmark$ & $\checkmark$ \\
\textbf{T1078} & \textbf{E.3.3} & $\checkmark$ & $\checkmark$ & $\checkmark$ & $\checkmark$ \\
\textbf{T1098} & \textbf{E.3.3} & $\checkmark$ & $\checkmark$ & $\checkmark$ & $\checkmark$ \\
\textbf{T1190} & \textbf{E.3.3} & $\checkmark$ & $\checkmark$ & $\checkmark$ & $\checkmark$ \\
\textbf{T1199} & \textbf{E.3.3} & $\checkmark$ & $\checkmark$ & $\checkmark$ & $\checkmark$ \\
\textbf{T1548} & \textbf{E.3.3} & $\checkmark$ & $\checkmark$ & $\checkmark$ & $\checkmark$ \\
\textbf{T1552} & \textbf{E.3.3} & $\checkmark$ & $\checkmark$ & $\checkmark$ & $\checkmark$ \\
\textbf{T1555} & \textbf{E.3.3} & $\checkmark$ & $\checkmark$ & $\checkmark$ & $\checkmark$ \\
\textbf{T1606} & \textbf{E.3.3} & $\checkmark$ & $\checkmark$ & $\checkmark$ & $\checkmark$ \\
\textbf{T1484} & \textbf{E.3.6} & $\checkmark$ & $\checkmark$ & $\checkmark$ & $\checkmark$ \\
\textbf{T1041} & \textbf{E.3.7} & $\checkmark$ & $\checkmark$ & $\checkmark$ & $\checkmark$ \\
\textbf{T1046} & \textbf{E.3.7} & $\checkmark$ & $\checkmark$ & $\checkmark$ & $\checkmark$ \\
\textbf{T1071} & \textbf{E.3.7} & $\checkmark$ & $\checkmark$ & $\checkmark$ & $\checkmark$ \\
\textbf{T1190} & \textbf{E.3.7} & $\checkmark$ & $\checkmark$ & $\checkmark$ & $\checkmark$ \\
\textbf{T1025} & \textbf{G.2.6} & $\checkmark$ & $\checkmark$ & $\checkmark$ & $\checkmark$ \\
\textbf{T1213} & \textbf{G.2.6} & $\checkmark$ & $\checkmark$ & $\checkmark$ & $\checkmark$ \\
\textbf{T1552} & \textbf{G.2.6} & $\checkmark$ & $\checkmark$ & $\checkmark$ & $\checkmark$ \\
\textbf{T1078} & \textbf{G.4.1} & $\checkmark$ & $\checkmark$ & $\checkmark$ & $\checkmark$ \\
\textbf{T1195} & \textbf{P.4.3} & $\checkmark$ & $\checkmark$ & $\checkmark$ & $\checkmark$ \\
\textbf{T1190} & \textbf{P.5.2} & $\checkmark$ & $\checkmark$ & $\checkmark$ & $\checkmark$ \\
\textbf{T1195} & \textbf{P.5.2} & $\checkmark$ & $\checkmark$ & $\checkmark$ & $\checkmark$ \\
\textit{T1190} & \textit{D.1.7} &  & $\checkmark$ &  & $\checkmark$ \\
\textit{T1195} & \textit{D.1.7} &  & $\checkmark$ &  & $\checkmark$ \\
\textit{T1199} & \textit{D.1.7} &  & $\checkmark$ &  & $\checkmark$ \\
\textit{T1027} & \textit{E.3.11} &  & $\checkmark$ & $\checkmark$ &  \\
\textit{T1036} & \textit{E.3.11} &  & $\checkmark$ & $\checkmark$ &  \\
\textit{T1080} & \textit{E.3.11} &  & $\checkmark$ & $\checkmark$ &  \\
\textit{T1098} & \textit{E.3.11} &  & $\checkmark$ &  & $\checkmark$ \\
\textit{T1190} & \textit{E.3.11} &  & $\checkmark$ &  & $\checkmark$ \\
\textit{T1195} & \textit{E.3.11} &  & $\checkmark$ &  & $\checkmark$ \\
\textit{T1199} & \textit{E.3.11} &  & $\checkmark$ &  & $\checkmark$ \\
\textit{T1221} & \textit{E.3.11} &  & $\checkmark$ & $\checkmark$ &  \\
\textit{T1564} & \textit{E.3.11} &  & $\checkmark$ & $\checkmark$ &  \\
\textit{T1584} & \textit{E.3.11} &  & $\checkmark$ &  & $\checkmark$ \\
\textit{T1608} & \textit{E.3.11} &  & $\checkmark$ &  & $\checkmark$ \\
\textit{T1195} & \textit{G.5.5} &  & $\checkmark$ &  & $\checkmark$ \\
\textit{T1199} & \textit{G.5.5} &  & $\checkmark$ &  & $\checkmark$ \\
\end{longtable}

\section{AUS mitigating tasks}
\begin{tabularx}{\linewidth}{rllllllX}
\toprule
\textbf{Score} & \textbf{\Gls{safeguard}} & \textbf{\gls{solarwinds}} & \textbf{\gls{log4j}} & \textbf{\gls{xz}} & \textbf{Level} & \textbf{Gap} & \textbf{\gls{attck} \gls{attack}}\\
\midrule
90 & E.3.3 & $\checkmark$ & $\checkmark$ & $\checkmark$ & $\updownarrow$ & $\CIRCLE$ & T1484, T1606, T1003, T1185, T1548, T1190, T1021, T1199, T1556, T1068, T1047, T1218, T1213, T1552, T1078, T1072, T1098, T1059, T1134, T1036, T1110, T1558, T1210, T1565, T1555, T1569, T1505, T1553, T1649, T1550 \\
87 & D.2.1 & $\checkmark$ & $\checkmark$ & $\checkmark$ & $\updownarrow$ & $\RIGHTcircle$ & T1484, T1606, T1041, T1003, T1548, T1021, T1190, T1556, T1068, T1568, T1218, T1213, T1552, T1078, T1573, T1105, T1203, T1059, T1027, T1071, T1036, T1195, T1562, T1558, T1210, T1555, T1543, T1550, T1070 \\
75 & E.3.7 & $\checkmark$ & $\checkmark$ & $\checkmark$ & $\updownarrow$ & $\RIGHTcircle$ & T1571, T1041, T1048, T1090, T1021, T1190, T1199, T1102, T1047, T1568, T1133, T1552, T1046, T1573, T1095, T1072, T1572, T1105, T1203, T1659, T1071, T1210, T1001, T1219, T1567 \\
33 & E.3.6 & $\checkmark$ & $\checkmark$ & $\checkmark$ & $\updownarrow$ & $\CIRCLE$ & T1484, T1098, T1059, T1114, T1552, T1548, T1505, T1562, T1543, T1556, T1550 \\
27 & E.3.11 & $\checkmark$ & $\checkmark$ & $\checkmark$ & $\updownarrow$ & $\Circle$ & T1098, T1027, T1036, T1195, T1190, T1199, T1584, T1564, T1608 \\
18 & P.5.2 & $\checkmark$ & $\checkmark$ & $\checkmark$ & $\updownarrow$ & $\LEFTcircle$ & T1210, T1195, T1190, T1553, T1068, T1072 \\
18 & P.2.1 & $\checkmark$ & $\checkmark$ & $\checkmark$ & $\updownarrow$ & $\CIRCLE$ & T1210, T1195, T1199, T1505, T1046, T1553 \\
15 & E.3.4 & $\checkmark$ & $\checkmark$ & $\checkmark$ & $\updownarrow$ & $\CIRCLE$ & T1659, T1565, T1213, T1114, T1552 \\
12 & G.2.6 & $\checkmark$ & $\checkmark$ &  & $\updownarrow$ & $\CIRCLE$ & T1041, T1565, T1048, T1213, T1552, T1005 \\
12 & D.1.2 & $\checkmark$ & $\checkmark$ & $\checkmark$ & $\updownarrow$ & $\RIGHTcircle$ & T1195, T1190, T1562, T1553 \\
9 & P.4.3 & $\checkmark$ & $\checkmark$ & $\checkmark$ & $\updownarrow$ & $\CIRCLE$ & T1210, T1195, T1190 \\
9 & G.4.3 & $\checkmark$ & $\checkmark$ & $\checkmark$ & $\updownarrow$ & $\CIRCLE$ & T1210, T1068, T1195 \\
9 & G.4.1 & $\checkmark$ & $\checkmark$ & $\checkmark$ & $\updownarrow$ & $\CIRCLE$ & T1078, T1195, T1190 \\
9 & E.3.2 & $\checkmark$ & $\checkmark$ & $\checkmark$ & $\updownarrow$ & $\CIRCLE$ & T1078, T1072, T1190 \\
9 & E.2.5 & $\checkmark$ & $\checkmark$ & $\checkmark$ & $\updownarrow$ & $\CIRCLE$ & T1068, T1072, T1190 \\
9 & D.1.7 & $\checkmark$ & $\checkmark$ & $\checkmark$ & $\updownarrow$ & $\Circle$ & T1195, T1190, T1199 \\
9 & P.4.5 & $\checkmark$ & $\checkmark$ & $\checkmark$ & $\updownarrow$ & $\CIRCLE$ & T1195, T1505, T1553 \\
9 & G.5.1 & $\checkmark$ & $\checkmark$ & $\checkmark$ & $\updownarrow$ & $\CIRCLE$ & T1195, T1190, T1199 \\
6 & D.1.4 & $\checkmark$ & $\checkmark$ & $\checkmark$ & $\updownarrow$ & $\RIGHTcircle$ & T1195, T1553 \\
6 & G.5.5 & $\checkmark$ & $\checkmark$ & $\checkmark$ & $\updownarrow$ & $\Circle$ & T1195, T1199 \\
6 & E.3.8 & $\checkmark$ & $\checkmark$ &  & $\downarrow$ & $\CIRCLE$ & T1078, T1552, T1558 \\
6 & D.1.3 & $\checkmark$ & $\checkmark$ & $\checkmark$ & $\updownarrow$ & $\CIRCLE$ & T1195, T1190 \\
4 & E.1.3 & $\checkmark$ & $\checkmark$ &  & $\downarrow$ & $\CIRCLE$ & T1078, T1098 \\
3 & P.5.1 & $\checkmark$ & $\checkmark$ & $\checkmark$ & $\updownarrow$ & $\CIRCLE$ & T1195 \\
3 & P.4.1 & $\checkmark$ & $\checkmark$ & $\checkmark$ & $\updownarrow$ & $\CIRCLE$ & T1195 \\
3 & P.3.4 & $\checkmark$ & $\checkmark$ & $\checkmark$ & $\updownarrow$ & $\CIRCLE$ & T1195 \\
3 & G.1.4 & $\checkmark$ & $\checkmark$ & $\checkmark$ & $\updownarrow$ & $\CIRCLE$ & T1195 \\
3 & G.3.1 & $\checkmark$ & $\checkmark$ & $\checkmark$ & $\updownarrow$ & $\CIRCLE$ & T1195 \\
3 & D.1.6 & $\checkmark$ & $\checkmark$ & $\checkmark$ & $\updownarrow$ & $\CIRCLE$ & T1195 \\
3 & G.1.5 & $\checkmark$ & $\checkmark$ & $\checkmark$ & $\updownarrow$ & $\CIRCLE$ & T1195 \\
2 & G.5.2 & $\checkmark$ &  & $\checkmark$ & $\uparrow$ & $\CIRCLE$ & T1072 \\
2 & G.2.5 & $\checkmark$ & $\checkmark$ &  & $\downarrow$ & $\CIRCLE$ & T1190 \\
2 & E.3.9 & $\checkmark$ & $\checkmark$ &  & $\downarrow$ & $\CIRCLE$ & T1078 \\
2 & P.4.4 & $\checkmark$ & $\checkmark$ &  & $\downarrow$ & $\CIRCLE$ & T1190 \\
2 & E.1.5 & $\checkmark$ &  & $\checkmark$ & $\uparrow$ & $\CIRCLE$ & T1072 \\
2 & E.2.4 & $\checkmark$ & $\checkmark$ &  & $\downarrow$ & $\CIRCLE$ & T1190 \\
2 & G.3.4 & $\checkmark$ &  &  & $\downarrow$ & $\CIRCLE$ & T1606, T1550 \\
\bottomrule
\multicolumn{8}{c}{$\checkmark$ : Mitigates \gls{aus}. Mitigation level: $\uparrow$ = Dependency, $\downarrow$ = Dependent, $\updownarrow$ = Both. Gap: $\Circle$ = Full, $\LEFTcircle$ = Major, $\RIGHTcircle$ = Minor, $\CIRCLE$ = None.}\\
\end{tabularx}

\end{document}